\documentstyle[11pt,epsf,fullpage]{article}

\title{{\bf Event-by-Event Analysis \\ and the \\ Central Limit Theorem}}

\author{{\bf T. A. Trainor}\\ \\
{\em Nuclear Physics Laboratory 354290}\\
{\em University of Washington}\\
{\em Seattle, WA 98195}\\
{\em trainor@hausdorf.npl.washington.edu}}
			       
\newcommand{\xfig}[1]{\begin{center}
\mbox{\epsffile{#1}}
\end{center}}

\newcommand{\pawplot}[1]{\begin{center}
\mbox{\epsffile[70 180 520 640]{#1}}
\end{center}}
\begin{document}
\maketitle

\begin{abstract}
Event-by-event analysis of heavy-ion collision events is an important tool for the study of the QCD phase boundary and formation of a quark-gluon plasma. A universal feature of phase boundaries is the appearance of increased fluctuations of conserved measures as manifested by excess measure variance compared to a reference.  In this paper I consider a particular aspect of EbyE analysis emphasizing global-variables variance comparisons and the central limit theorem. I find that the central limit theorem is, in a broader interpretation, a statement about the scale invariance of total variance for a measure distribution, which in turn relates to the scale-dependent symmetry properties of the distribution.. I further generalize this concept to the relationship between the scale dependence of a covariance matrix for all conserved measures defined on a dynamical system and a matrix of covariance integrals defined on two-point measure spaces, which points the way to a detailed description of the symmetry dynamics of a complex measure system. Finally, I relate this generalized description to several recently proposed or completed event-by-event analyses.
\end{abstract}

\section{Introduction}

With the advent of high-multiplicity heavy-ion collisions at the CERN SPS
and RHIC there is the possibility to extract statistically significant
dynamical information from individual collision events.  By determining the
degree of correlation of the multiparticle final state for individual
events within a large ensemble of events one can probe the dynamical
history of collisions and the QCD phase boundary.  This process is called
event-by-event (EbyE) analysis.

A central issue for EbyE analysis is the question whether or not collision
events with similar initial conditions experience the same dynamical
history, excepting finite-number fluctuations and known hadronic variance
sources.  If similarly prepared events differ significantly from one
another beyond expectation, then analysis must proceed to categorize
events into dynamical classes or on a continuum, and to provide a physics
interpretation for nontrivial variations.

Based on very general arguments and the observed behavior of normal bulk 
matter near phase boundaries we expect to observe significant changes in fluctuations near the QCD phase boundary. The amplitude and 
character of these changes and their observability given finite 
collision systems and non-equilibrium collision trajectories are major 
questions for EbyE physics. Experimentally, the problem calls for 
extracting all available information from each event by  a complete
topological analysis in order to obtain maximum sensitivity to  dynamical
fluctuations.

Various measures can be used to categorize events by their `morphology.'  
The common element of all measures is the correlation
content of the multiparticle final state.  Correlation measures can be
loosely grouped into {\em global measures}, some of which have
counterparts in thermodynamics, and {\em scale-local measures}, which
explicitly determine the scale dependence of correlations.
Given an event-wise measure we also require a corresponding {\em
comparison}\/ measure to enable quantitative event comparisons.

In this paper I emphasize EbyE analysis techniques which relate to
variances and second moments.  These global statistical measures are
directly related to two-point correlation measures defined in various pair
spaces.  Higher moments are related to general $q$-point correlations and
are more efficiently dealt with by a generalized system of {\em scale-local} EbyE analysis which will be treated in subsequent papers.  The emphasis on
variances and two-point correlations in this paper is both timely and
provides a reasonably accessible introductory treatment of EbyE analysis.

The paper describes the general types of EbyE analysis, the available
system of correlation measures, the system of ensembles and its
relationship to topological partition systems, considers maximally
symmetric systems and the statistical model as archetypal correlation
reference systems, the relationship between fluctuations and correlations, introduces the central limit theorem as an archetype for variance comparison,
considers a fluctuation comparison measure ($\Phi_{p_t}$) as a difference
between $rms$ fluctuation amplitudes, analyzes the structure of
$\Phi_{p_t}$ as a specific variance comparison measure, introduces the
concept of scale-dependent covariance matrices as a generalized measure
system, relates scale-dependent covariance to a system of two-point
correlation integrals, and considers the novel connections between scale invariance
and the central limit theorem which arise from this work.

\section{Correlation Measures and EbyE Analysis}

Correlation measures provide the basic analytical infrastructure for EbyE
analysis.  A complete characterization of the correlation content of the
multiparticle final state should exhaust all significant information in the
data.  Comparison of event-wise correlations in a data system with those
of a reference system (which may invoke `equilibrium' or an assumption of
maximum symmetry) offers a precise means to detect deviations from a
model or ensemble average.  Significant deviations provide a basis for discovering new physics.  One then attempts to understand the physical origin of these deviations, leading to a revised model as part of an oscillatory (but
hopefully converging) scientific process.

\subsection{EbyE analysis types}

\begin{figure}[th]

\begin{tabular}{cc}

\begin{minipage}{.57\linewidth}
\epsfysize .35\textwidth
\xfig{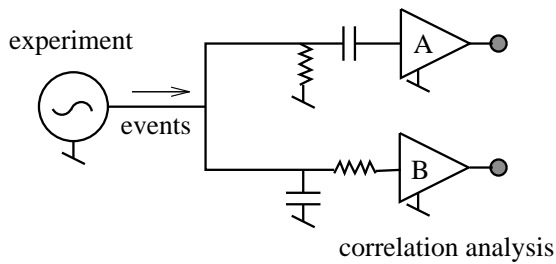}  
\end{minipage} &

\begin{minipage}{.37\linewidth}
\epsfysize 1\textwidth
\caption{Signal processing analog of EbyE analysis showing AC ($A$) and DC 
($B$) aspects. Type $A$ analysis emphasizes the differences in correlation 
content among individual events. Type $B$ analysis emphasizes the 
correlation content common to an ensemble of events relative to a reference 
population.\label{filter2}}
\end{minipage}

\end{tabular}


\end{figure}

We can distinguish two types of event-by-event analysis as represented in
Fig.  \ref{filter2}.  In analysis type $A$ each event is fully
characterized for information content and recorded in an `event spectrum,'
a space designed to represent differing event morphologies in an efficient
manner.  This could be as simple as a 1-D space recording the event-wise
slope parameter of $ m_t$ distributions or the event-wise $<\!p_t\!>$. 
Or, it could be a complex scale-dependent representation of correlations
in ($\eta, \phi$) distributions for a minijet analysis. Through the
intermediary of event-spectrum distributions object and reference event
spectra are compared for nonstatistical differences in event  morphology.
In this analysis we are usually not interested in the `DC'  aspects of
correlation common to every event, but only in those `AC'  aspects that
change from one event to another.

Type A analysis generally requires large acceptance and large event
multiplicity.  One attempts to characterize all correlation aspects of each
collision event over most of phase space, at least implicitly using
$q$-particle correlations, where $q$ may range up to a significant fraction
of the event multiplicity.  Questions of interest include large-scale
variations in collision dynamics (dynamical fluctuations) such as flow 
fluctuations, baryon stopping fluctuations, jets, departures from chemical 
and thermal equilibration and chiral symmetry variations. The identity of 
individual events is preserved for essentially all aspects of this analysis.

In analysis type $B$ elements of each event are accumulated in an
inclusive {\em sibling}\/ space ({\em e.g.,} a pion pair space for HBT
analysis or an azimuthal space with registered event planes for flow
analysis) and compared to a mixed-element space (no siblings by
construction, hence minimally correlated) to determine differential
inclusive symmetry properties of the event ensemble.  Here we are
interested in the `DC' aspects of correlation  which are true for almost
every event, but which may deviate from a model expectation. The identity
of individual events is typically discarded  in type-$B$ analysis.

If statistical measures (such as moments) are extracted from an event 
spectrum obtained in a type-$A$ analysis in such a way that event 
identities are lost then the result is effectively a type-$B$ analysis, 
even though the statistics were obtained through a type-$A$ analysis. For 
instance, the moments of a $<\!p_t\!>$ distribution form the basis for a 
type-$B$ analysis, although the $<\!p_t\!>$ distribution itself is the 
event spectrum for a type-$A$ analysis. The variance of a $<\!p_t\!>$ 
distribution contains information which is closely related to that of an 
inclusive pair-correlation analysis. As we shall see however, this is a 
complex relationship. In contrast, an analysis in which features of an 
event spectrum are used to select special event classes is definitely a 
type-$A$ analysis. Special event classes can then be used as the basis
for  conventional {\em inclusive}\/ analyses ({\em e.g.,} particle ratios
or  baryon stopping) which would not be feasible on a single-event basis
due to limited statistical power but may be essential to interpret the
special event  class.

For type-$B$ analysis large acceptance and large event multiplicity are
still highly desirable. For N-N {\em vs} A-A comparisons type-$B$ analysis may be the
only practical study method due to the low multiplicity of N-N events. Comprehensive analysis of two-point densities over a large acceptance requires {\em simultaneous} measurement over the entire two-point space to minimize systematic error. As
we shall see, in-depth EbyE analysis is often statistics limited and
depends on access to the largest possible multiparticle phase space,
practical conditions that require the largest possible detector
acceptance.  EbyE analysis based on variance comparisons as discussed in
this paper represents only the most elementary initial approach.  More
complex extensions to arbitrary $q$-tuple analysis and scale-local
analysis of higher-dimensional measure distributions will extend the initial
studies of simple 1-D distributions described here.

\subsection{Correlation measures} \label{sca}

The basic tools of EbyE analysis are  correlation measures.  Each
multiparticle event distribution has a certain information content
relative to a reference.  In principle, all significant event information
can be extracted to an efficient system of correlation measures by a
data-compression process. If a distribution nears maximum symmetry within
constraints the event information may be represented by a very simple
measure system.  We can divide correlation measures roughly into global
measures, which are effectively integrals over some scale interval, and
scale-dependent or scale-local measures.  A central aspect of EbyE
analysis is comparison between object distributions and references.  Thus,
every correlation measure requires an associated comparison measure.  A
summary of correlation measures is presented in Table \ref{table}.

\begin{table}[th]

\begin{tabular}{cc}
\begin{minipage}{.64\linewidth}
\epsfysize 1\textwidth

\begin{center}

\begin{tabular}{|c||c|} 
\hline
 measure  &  comparison measure  \\ \hline\hline
\multicolumn{2}{|c|}{ \bf global measures} \\ \hline\hline
 moments  &  central-limit  tests  \\ \hline
 $W_m(D) = <\!x^m\!>^{1/m}$  &  $\Phi_x(D_1,D_2,m)$  \\ \hline\hline
 model distributions  &  {\em GoF}\/ criteria  \\ \hline
  $f(x,\alpha)$ &  $\chi^2\{f(x,\alpha),D\}$  \\ \hline\hline
\multicolumn{2}{|c|}{ \bf scale-dependent measures} \\ \hline\hline
 factorial moments  &  normed factorial moments  \\ \hline
  $\mu_{(q)}(D,\delta x)$ &  $F_q(D_1,D_2,\delta x)$  \\ \hline\hline
\multicolumn{2}{|c|}{ SCA} \\ \hline
 $S_q(D,\delta x)$ entropy  &  $I_q(D_1,D_2,\delta x)$ information  \\ 
\hline
  $d_q(D,\delta x)$ dimension &  $\Delta d_q(D_1,D_2,\delta x)$ dim. 
transp.  \\ \hline
 $V_q(D,\delta x)$ volume &  $\rho_q(D_1,D_2,\delta x)$ density  \\ \hline
 \end{tabular}

\end{center}

\end{minipage} &

\begin{minipage}{.3\linewidth}
\epsfysize 1\textwidth
\caption{ Correlation measures and associated comparison measures. $D$ is a 
distribution or data set, ($D_1,D_2$) is an object/reference pair of 
distributions or data sets, $\alpha$ is a set of model parameters, $GoF$ is 
goodness-of-fit, $q$ is a $q$-tuple index, $\delta x$ is a scale (bin size) 
parameter. \label{table}}

\end{minipage}

\end{tabular}

\end{table}

{\em Global measures}\/ include distribution moments and model parameters.  The respective comparison measures are moment
differences ({\em e.g.,} central-limit tests) and goodness-of-fit ($GoF$)
measures ($\chi^2$ or likelyhood).  The reference may be represented
either by a set of moments or by a restricted region in the parameter
space of a model distribution.  Event-wise information then consists of
deviations from the reference represented by moment differences or
parameter deviations.  As we shall see, global measures are effectively
based on scale integrals, although this may not be obvious from typical
usage.  Integral measures are more appropriate in the case of low
statistical power (low event multiplicity) and small correlation content,
when only one or a few significant parameters can be extracted from an
event.

{\em Scale-dependent measures}\/ become more important with increasing event 
multiplicity and/or correlation amplitude. It is then possible for an event to carry more information, and one can determine the scale and space dependence of correlations. A system of measures based on the scale-dependent correlation integral has been 
developed in two closely-related manifestations. Factorial moment analysis 
has a substantial history in multiparticle analysis \cite{bia} and is 
based on the factorial moment as scale-dependent measure and the normalized factorial moment or moment ratio as the comparison measure. Scaled 
correlation analysis (SCA) is based on a system of generalized entropies 
\cite{re1,re2} extended to scale-local form. The three closely-related 
topological measures entropy, dimension and volume are indexed by $q$-tuple number and scale. The corresponding comparison measures are information, 
dimension transport and density. In some applications model distributions such as the negative-binomial distribution have been employed in scale-dependent analyses \cite{tan,chi}. It is possible to make a connection between
scale-dependent model parameters  and SCA measures.

\section{Ensembles} \label{ensemb}

Many-body dynamics, equilibration and fluctuations -- issues central to
EbyE analysis -- can be studied with an ensemble approach. We
encounter ensemble and partition concepts repeatedly in this paper.  An
ensemble in the sense of statistical mechanics is a set of nominally
identical dynamical systems (events) prepared with the same constraints but
otherwise uncorrelated.  Under Gibbs' ergodic hypothesis the phase-space
distribution of system states of the ensemble at any time should reflect
the phase-space history of an individual system state over a long
interval.  For nearly symmetric systems this hypothesis provides an
essential reference.

\begin{figure}[th]
\epsfxsize .65\textwidth 
\xfig{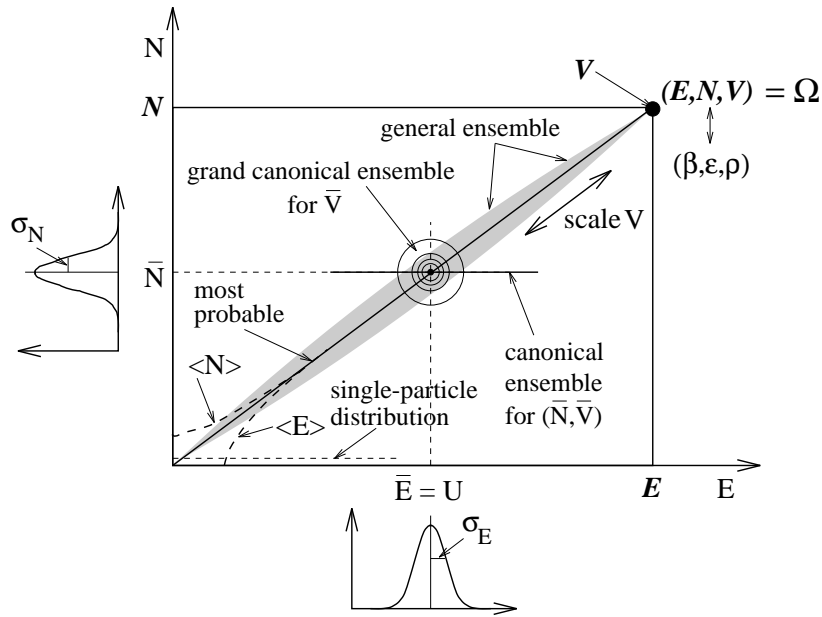}
\caption{Ensembles: An isolated volume $\Omega$ with extensive properties 
(${\bf \cal E,N} \cdots$) is defined. Various subsets of the class of all 
possible partitions of $\Omega$ can be identified as ensembles. Contents of partition elements are represented as points in the single-point measure space $G_1(\vec{m})$.  Points representing the general ensemble are localized near the main diagonal of this space. The projection of $G_1(\vec{m})$ for the GE onto the $G_1(E,N)$ plane is shown as the shaded region in the figure. The grand canonical ensemble is represented by a slice through this distribution at fixed volume $\overline{V}$. The result is a localized distribution (concentric circles) centered at ($\overline{E},\overline{N}$) with variances $\sigma^2_E$ and $\sigma^2_N$. \label{canonical}}
\end{figure}

A generalized ensemble can be formed from an arbitrarily large bounded
volume $\Omega$ containing a measure system in `statistical equilibrium'
(maximally symmetric within global constraints).  $\Omega$ is invested
with a set of conserved global extensive quantities (additive measures)
such as total energy ${\bf \cal{E}}$, total volume ${\bf \cal{V}}$ and
total object number ${\bf \cal{N}}$.  This list extends to all quantites
conserved by a given dynamical system and is represented below by $\vec{m} = \{m_i\}$.  The global volume is partitioned
arbitrarily into elementary volumes $V_e$.  Each volume element $V_e$
contains local samples of extensive quantities such as energy $E_e$ and
number $N_e$.  Each sample set $(E,N)_e$ provides local estimators
$(E/N,E/V,N/V)_e$ for global intensive quantities $(T,\epsilon,\rho)$
representing the global constraints ${\bf (\cal{E},\cal{V},\cal{N})}$. 
The (infinite) set of elements from all possible partitions of $\Omega$ I
designate the {\em general ensemble}\/ (GE).  This description of a general ensemble adheres closely to the language of Borel
measure theory so that topological analysis methods can more easily be
connected to the statistical study of dynamical systems.  A system of
ensembles is illustrated in Fig.  \ref{canonical}.

The measure contents of GE partition elements can be represented in a single-point measure space $G_1(\vec{m})$ (cf Secs. \ref{pg} and \ref{covmat}). The measure space in Fig. \ref{canonical} is a subspace $G_1(E,N)$. For a minimally correlated system the GE measure distribution is localized near the main diagonal of $G_1(\vec{m})$.  The degree of localization
(fluctuation amplitude or variance) depends on the {\em effective} total number of
dynamical objects ${\bf \cal{N}}$ in the space $\Omega$ and the fractional
volume or scale of typical ensemble elements measured by position along
the main diagonal. Note that fluctuation amplitudes decrease (are
`suppressed') as this partition scale approaches the total volume ${\bf
\cal{V}}$ of the space $\Omega$.  The distribution becomes sensitive to
the global constraints or boundedness of the dynamical system.

The classical microcanonical, canonical and grand canonical ensembles are
conventional subsets of the GE.  The grand canonical ensemble (GCE)
consists of all partition elements with some fixed volume $V$ (system open
to particle and energy exchange).  The canonical ensemble (CE) is a special
case of the GCE in which object number $N$ is held fixed (system open to
energy exchange).  The microcanonical ensemble is a further special case in
which ($E,N,V$) are all held fixed (closed statistical system).  $\Omega$
itself can be seen as an element of a microcanonical ensemble for a larger
system. More generally, given an arbitrary measure system defined on a partitioned multidimensional space, one can define a hierarchy of subensembles of successively lower dimensionality corresponding to increasing degrees of constraint on partition elements and measures.

For the purpose of EbyE analysis of nuclear collisions the GCE for an equilibrated system can
serve as a minimally correlated (maximally symmetric) reference for the
transverse phase space of an event ensemble.  One asks whether the
stopping process (source of total $P_t$) for a fixed macroscopic collision
geometry is the same for each event, excepting finite-number fluctuations,
and whether the `stopped' energy is fully equilibrated among transverse
degrees of freedom. 

To answer these questions one must compare observed measure distributions
for an event ensemble to a suitable reference such as the GCE for an
equilibrated system. Multiplicity can be compared to a Poisson,
quantum-statistical or hadronic resonance-gas reference.  Mean transverse momentum per
particle can be compared to an inclusive-spectrum reference, and the
linear correlations between these variable pairs can also be compared to
an appropriate covariance reference, including effects of hadronic correlations.  Agreement between data and the GCE might indicate that in each event $N$ {\em independent}\/ particle momenta
are sampled from a {\em fixed}\/ inclusive momentum distribution (parent),
consistent with independent particle production from an invariant source. 
Such a comparison test is closely related to the central limit theorem
(CLT) and represents the subject of much of this paper.

\section{Equilibration and the Statistical Model}

By the process of equilibration a many-body system in some arbitrary
initial state approaches a state of maximum symmetry.  To study this
process with sensitivity one needs a well-defined reference and
optimized comparison measures.  The simplest reference corresponds to an
assumption of maximum symmetry within a constraint system, which leads to a
`statistical-model' description.  Observed deviations from a statistical
model may then indicate that an equilibration process from a less symmetric
initial state has terminated before achievement of maximum symmetry or that the true constraint system is incorrectly modeled.

The specific term `thermalization' connotes equilibration of
energy in contrast to a more general situation in which various
quantities such as charge, color, flavor, spin and baryon number
may undergo transport in the approach to maximum symmetry.  Equilibration is therefore the preferable term.  Equilibrium (zero {\em net} average transport of conserved quantities) and maximum symmetry are essentially synonymous asymptotic conditions.

\subsection{The statistical model}

To establish a robust reference
system for detecting possibly small deviations which may convey new physics we apply a hypothesis of maximum symmetry (Occam's Razor, maximum entropy): we adopt the `simplest' hypothesis consistent with global constraints and prior information.  The corresponding many-body system is said to be in statistical equilibrium and its mean-value behavior can be described by a {\em statistical model}\/ (SM).

The statistical model has been applied to simple systems
including four-momentum and strangeness as additive measures
\cite{be1,be2,be3}. A dynamical system in some initial state (possibly
partonic) is partitioned at large scale into elements  (clusters, hadronic
`fireballs,' jets, flow elements) having arbitrary mean momenta.  Small-scale
system homogeneity is assumed (all clusters have the same intensive 
properties T and $\gamma_s$ at hadronization). The large-scale mean
momenta of clusters comprise an unthermalized collective flow field which
is  assumed to play no role in determining particle species abundances.
Abundances and ratios are instead assumed to be determined by thermalized
energy residing at small scale, with no overlap between these two scale
regions.

The partition function for an event composed of individual
clusters is a simple product of cluster partition functions with
no cross terms, implying linear independence of clusters.  Each cluster partition function in turn treats elements (particles) as members of an uncorrelated gas (except for quantum correlations).  The treatment is however `canonical' in that some extensive quantities are event-wise conserved, thus introducing a degree
of correlation. As we shall see, these assumption of homogeneity,
linearity and independence are also preconditions for the central
limit  theorem.

With the exception of the large-scale flow field the system is assumed to
be scale-invariant and maximally symmetric (minimally correlated).
Therefore, the system state is determined only by global
constraints. Global fitting parameters are T, V and $\lambda_s$ (quark
strangeness suppression factor) or $\gamma_s$ (hadronic strangeness
saturation factor or fugacity).  V is the single extensive quantity and
should depend on the collision system.  T may depend only on the
fundamental process of hadronization in elementary systems.  An asymptotic
(maximum) T value may be observed for small systems in which the hadronic
cascade is quickly arrested.  In the limit of large system volume and/or
particle number this formalism may become consistent  with a
grand-canonical approach.

The statistical model provides  a satisfactory {\em general} description of
multiplicity  distributions (revealing primary-hadron or clan
correlations) and hadronic  abundances/ratios for small and large
collision systems over a broad collision-energy interval.
An SM treatment of hadron multiplicity distributions for light systems
\cite{be1} results in two-tiered distributions, with primary hadrons
following a nearly Poisson distribution and secondary hadrons distributed
according to a negative binomial distribution (NBD) consistent with the
number of primary hadrons predicted by the SM. SM treatments of hadron abundances or abundance ratios \cite{be2,be3} and strangeness suppression \cite{be4} in a variety of collision systems lead to good general agreement with experimental observations in terms of a single universal hadronization temperature $T \approx 180$ MeV and volumes $V$ consistent with specific collision  systems from $e^+$-$e^-$ to the heaviest HI collisions. Although the SM provides a good general summary of hadronic abundances interesting physics may reside in significant disagreements with this model. The SM provides an equilibration reference from which deviations should be the subject of further study.

\subsection{Equilibration}

\begin{figure}[th]
\begin{tabular}{cc}
\begin{minipage}{.60\linewidth}
\epsfxsize .8\textwidth
\pawplot{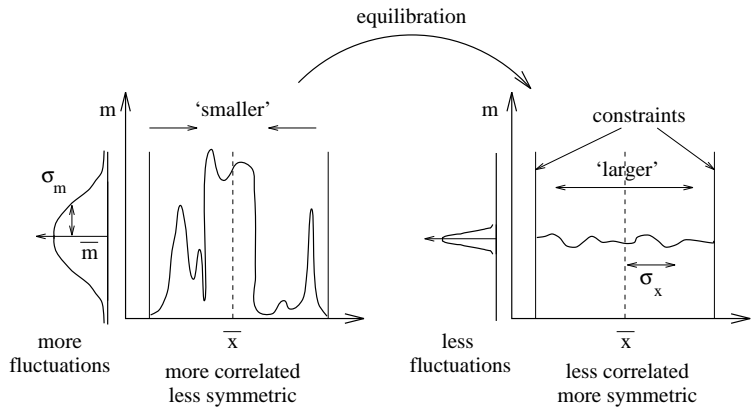}  
\end{minipage} &
\begin{minipage}{.34\linewidth}
\epsfysize 1\textwidth
\caption{Variation of fluctuations, correlations and symmetry with advancing equilibration. Decreasing correlation corresponds to increase of effective system size (entropy) within constraints, increased symmetry and decreased measure variation (fluctuations). Thus, determination of fluctuations and/or correlations of measure distributions informs us about the equilibration state of a dynamical system.\label{fluctcorr}}
\end{minipage}
\end{tabular}
\end{figure}

A statistical system appears from a macroscopic viewpoint to move
persistently toward  a state of maximum overall symmetry in an
equilibration process. What underlies this apparent
teleology is understood to be a probabilistic effect:  maximally symmetric
macroscopic systems are more likely to be realized in the random evolution
of a normal-system microscopic phase space. Symmetry increases with effective phase-space volume (entropy) and {\em
dimensionality}.  Increased correlation {\em decreases}\/ the effective
symmetry, volume and  dimensionality of a system and corresponds to a less
likely macroscopic state. Equilibration then represents a general decrease
of system correlation toward a maximally symmetric reference as illustrated in Fig. \ref{fluctcorr}, and can be followed quantitatively with differential correlation measures.

We associate different equilibration types with corresponding 
transported quantities, such as thermal (momentum) and chemical (flavor) 
equilibration. The relaxation rates of these processes depend on local 
cross sections and flux densities. Different equilibration processes may 
effectively begin and end (freeze out) in different space-time regions of 
the collision, and with finite transport rates the degree of equilibration is limited by the system space-time extension \cite{uli}. An equilibration reference should therefore be defined in terms of finite system size/duration to insure proper treatment of the scale dependence of system symmetries \cite{par}.
Concepts such as `local' and `global' equilibration relate to scale
dependence, of vacuum symmetries as well as energy/momentum
equilibration.  Color deconfinement (spatial color symmetry) may be an
established condition within cold hadrons. One is more interested in how
color and chiral symmetries are modified and {\em spatially extended}\/
during a HI collision, what is the varying geometry or dimensionality of
regions with modified symmetries and how does this dimensionality
propagate on scale.  Microscopic variations due to small-scale QCD effects
are not as interesting as large-scale deviations from a global thermal
model that are {\em unique} to large collision systems.  Comparisons between
large- and small-system deviations and the scale dependence of correlation
within a given system are therefore important.

\subsection{Cascades and scale dependence}

In conventional treatments of an equilibrated many-body system scale
invariance is  assumed, and no provision is made to deal with departures
from this  assumption.  If equilibrium is not achieved there has been no
quantitative thermodynamic  measure system available to deal with deviations. The
partition systems  usually applied represent very elementary versions of
Borel measure theory and  treat scaling issues, if at all, with a limited
two- or three-tiered  structure. Few-stage partition systems in HI
collisions represented by  jets, strings or primary hadrons combined with
a hadronic cascade cannot  describe more complex departures from
equilibrium expected near a phase boundary.

A classical thermodynamic  treatment invokes an extreme scale partition
into an `atomic' scale and the system boundary scale. Work can be done on
the system boundary  `adiabatically,' meaning that energy transferred to
the boundary cascades down to the atomic scale within a time small
compared to the large-scale  energy transfer. The details of the cascade
are not described. In thermodynamic language PdV `work' represents 
large-scale energy transfer. TdS `heat' represents small-scale energy 
transfer. There is no intermediate possibility. What is `adiabatic'
depends on transfer times at large scale  relative to transport times
(collision rates) at atomic scale in the  thermodynamic medium.

In contrast, scale dependence is an essential ingredient of a
comprehensive statistical treatment of many-body dynamics. A cascade
process is a hierarchical structure representing the bidirectional
propagation of correlation and symmetry on scale. Correlation
spontaneously propagates down scale, and equivalently symmetry 
(dimensionality) propagates upscale in the dominant trend, but with local
deviations. In an open system a symmetry increase at larger scale may be
implemented `opportunistically' by an accompanying symmetry reduction at
smaller scale ({\em e.g.,} turbulence vortices, biological structures), a process sometimes referred to as `spontaneous' symmetry breaking.   

A falling stone collides with a pool of water both at its surface, depositing energy in surface waves, and along a path through its volume, creating a system of vortices. The energy in these large- and medium-scale correlations is transported to smaller scale by a cascade process termed collectively `dissipation' or `damping.' If the dissipation process or cascade is allowed to proceed indefinitely the stopped energy of the stone is finally transported to the smallest scale relevant to the problem, the molecular scale, and the system becomes maximally symmetric or thermalized. If the system is frozen or decoupled before this process is completed residual correlations over some scale interval may be observed. The analog in a HI collision is the interplay among momentum transfers at nuclear, nucleonic and partonic scales and subsequent evolution to final-state hadron momenta at freezeout.

The evolution of an expanding partonic medium into primary hadrons by a
coalescence process \cite{gel} involves local symmetry reduction
(coalescence) accompanying global symmetry increase achieved by the
expansion.  The decay of these primary hadrons to observed  secondaries is
a further cascade process, a symmetry increase in which detectable  correlations
nevertheless persist in the secondary hadron population unless the
hadronic cascade can proceed by continued rescattering to a state of
maximum symmetry -- an equilibrated hadron gas.

The terminology of cascade processes varies with context, as represented
by the many terms for localized dynamical objects (quasiparticles) in a nuclear collision: fireballs, primary hadrons, clans, prehadrons, clusters, participants, (wounded) nucleons, resonances, strings, leading partons -- resulting in various manifestations
of residual correlations in the hadronic final state such as jets,
minijets, Cronin effect and NBD. It is desirable to develop a simplified common
language which treats these various manifestations of symmetry variation
consistently.

\subsection{Equilibration tests}

Are chemical and thermal equilibrations achieved during a HI collision? 
Is the final state of an N-N, p-A or A-A collision the result of a
completed equilibration process?  Or are correlation remnants from the
early stages of the collision process still detectable in the multiparticle final state, and do they indicate evidence for QGP formation? Methods are required to distinguish between equilibrium and nonequilibrium models. These questions cannot be resolved by inclusive measurements. To detect and study possible {\em departures}\/ from equilibrium or maximum symmetry we require a differential correlation analysis system and a reference system analogous to the statistical model but applicable to {\em fluctuation}\/ analysis.  This paper focuses on measurement and interpretation of {\em variance differences} as one component of a global EbyE fluctuation analysis.

If remnants of a cascade structure are still present after
hadronic freezeout we may observe the corresponding residual correlations
as excess or deficient (co)variance (fluctuations) of global measures.  An
elementary equilibration test might therefore consist of comparing the
variances of event-wise measure-distribution statistics to a reference system.  A
nonequilibrated system might exhibit variances differing from finite-number
statistics and elementary hadronic processes. 

Ga\'{z}dzicki, Leonidov and Roland (GLR) \cite{glr} considered whether $p_t$ production in HI collisions is consistent with a
fully equilibrated system or retains evidence of initial-state or
pre-equilibrium scattering of projectile nucleons (ISS, Cronin effect). 
The initial state scattering (ISS) model predicts increased $p_t$
fluctuations because $p_t$ is generated by a {\em hierarchical process}\/
in which transverse random walk of projectile-nucleon $p_t$ (multiple
scattering) is followed by thermal generation of hadrons about the
resulting primary $p_t$ distribution. This relationship between
hierarchical structure and increased correlation is observed in studies of
NBD multiplicity distributions and clans in hadron production in small
collision systems \cite{be1}.   ISS should lead to a correlated $p_t$
population and corresponding fluctuation excess.  These fluctuations
should increase in amplitude with the number of possible nucleon
interactions (thus with nuclear size).

Both equilibrium-hydrodynamic and ISS models provide reasonable
descriptions of inclusive $p_t$ distributions.  To further distinguish
between the two alternatives GLR proposed to compare $p_t$ {\em
fluctuations}\/ predicted by the ISS model and a hydro model with
experimental data.  The $\Phi_{p_t}$ measure which they employed is said to
provide a quantitative basis for comparing fluctuations in HI
collisions relative to elementary interactions \cite{gnm}.

An additional source of $p_t$ fluctuations according to GLR is the
observed $N$-$<\!p_t\!>$ correlations in N-N collisions \cite{kaf}, some
degree of which may survive an A-A collision considered as a superposition
of elementary N-N events.  According to GLR surviving $N$-$<\!p_t\!>$
correlations would lead to departures from independent-particle emission
detectable with an appropriate measure such as $\Phi_{p_t}$ applied in a
systematic study of $N$-$<\!p_t\!>$ correlations in N-N, p-A and A-A. 
They also suggest that a substantial change in the nuclear medium in A-A
collisions could cause a different relationship among elementary N-N
collision subevents as indicated by a significant change in differential
variance.  They conclude that $\Phi_{p_t}$ should be sensitive both to
$N$-$<\!p_t\!>$ correlations and to ISS.

\section{Fluctuations and Correlations}

Fluctuations and correlations are complementary descriptive terms
reflecting the degree of structure in a measure distribution. The term
fluctuations may carry a connotation of time dependence, but the term
more generally refers to the {\em variability}\/ of a measure
distribution, whether temporal or spatial. Correlation refers to the {\em
localization}\/ or size reduction of a measure distribution on an
embedding space. Increased localization leads to increased variability of
the measure, possibly interpreted as fluctuations. The terms are sometimes used interchangeably.

Although fluctuations and correlations can be characterized by a complete
system of measure distribution moments this paper emphasizes second moments or
variances. The primary variance source in a typical `many-body' system is
the finite density of measure-carrying objects (degrees of freedom or
$DoF$). This concept emerges when a measure system accumulates significant correlation over restricted scale intervals (correlation onset
$\rightarrow $ `particles'), and only a small degree of correlation over
intervening scale intervals (see Sec. \ref{totvar}). As a simplification
one then focuses on the correlations {\em among}\/ $DoF$ at large scale, and ignores
their internal structure at smaller scale.

A reference density of independent $DoF$ corresponds to an expected or
basal variance in any physically observable quantity.  If the {\em
effective}\/ density of $DoF$ changes from a reference value observed
fluctuations may depart from expectation. A dramatic change in the number
of  $DoF$ can happen in a phase transition, in which scale-localized
correlations are created or destroyed by a change in system constraints. 

Changes in fluctuation amplitude can also happen if correlations among
$DoF$ change. Positive correlation reduces the {effective} number of $DoF$
leading to excess variance beyond expectations.  The opposite is true of anticorrelation.  One can say equivalently that the entropy associated with a class of $DoF$ is reduced through a coupling process, leading to a degree of freezeout of
the corresponding $DoF$ from an equipartition process.  Changes in
fluctuation amplitudes or variances can thus be seen as indicators of
increased or decreased {\em correlation among}\/ $DoF$ or as a change in the
{\em density of}\/ $DoF$, and can be used to probe the underlying Lagrangian
for a many-body system or the mechanism of a dynamical process.

Some correlations in a dynamical system are manifestations of incomplete
equilibration, the surviving remnants of an incomplete cascade following
initial large-scale departures from symmetry. Some correlations may be due
to quantum  statistics or other steady-state multibody interactions.  If
excess correlation has been detected beyond that attributable to known
sources the problem then lies in identifying the process that led to this
correlation and/or what reduced symmetry initiated it. From an EbyE
viewpoint deviations from {\em expected}\/ fluctuation amplitudes (either
more or less) which may signal unexpected physics or differences from a
reference model are of primary interest.

A central problem for HI physics is the nature of the QCD phase boundary.
The location of the boundary can be determined by inclusive measurements
\cite{uli}, but the detailed characteristics of the boundary are more directly probed
by EbyE analysis \cite{edw}. Fundamental to EbyE QCD studies is the
quantitative measurement of changes in QCD vacuum symmetries as
energy and baryon-number densities are varied.   The most accessible probe
of the QCD vacuum may be the multiparticle final state of an URHI
collision.  Near the phase boundary event-by-event dynamical fluctuations
should provide essential information on vacuum symmetry dynamics.

In the partonic regime the highly structured nonabelian gluon field plays
a substantial role \cite{shu}. At RHIC energies and above the incidence of
parton interactions with high momentum transfer should lead to jet
formation and a large minijet background in A-A collisions \cite{wag},
resulting in a highly correlated multiparticle final state at high $p_t$
due to hard processes.


At a phase boundary the density of momentum-carrying objects or degrees of
freedom ($DoF$) changes rapidly with certain state variables ({\em e.g.,}
temperature and net baryon density).  This means that in the neighborhood of
the boundary a bulk-matter system undergoes a sharp change in pressure and
energy density.  Fluctuations in extensive quantities exceed those
expected from finite-number statistics by an amount bounded by (but not determined by) a
corresponding linear response coefficient ({\em e.g.,} heat capacity)
\cite{par}.  To determine the detailed properties of the phase boundary it
is therefore important to search for and measure nonstatistical
fluctuations in dynamical quantities.

However, there are important differences between a phase boundary
in bulk  matter and {\em manifestations}\/ of a phase boundary
traversed by a small system during a collision, which manifestations may
be {\em barely detectable} in the multiparticle final  state \cite{par}.
Phase-boundary manifestations compete for observability with details of
the collision dynamics, the space-time finiteness of the collision  system
and the varying degrees of equilibration in different  space-time regions
of the collision. Potentially observable effects can be attenuated by
final-state hadronic interactions which move the system further toward
equilibrium, with consequent loss of information, and introduce distracting
hadronic correlations which are not of primary interest. There is a
critical need therefore to develop a sensitive, complete and interpretable system of
correlation measures and comparison measures for heavy-ion collisions.

\section{The Central Limit Theorem} \label{clt}

\begin{figure}[th]
\epsfxsize .5\textwidth 
\xfig{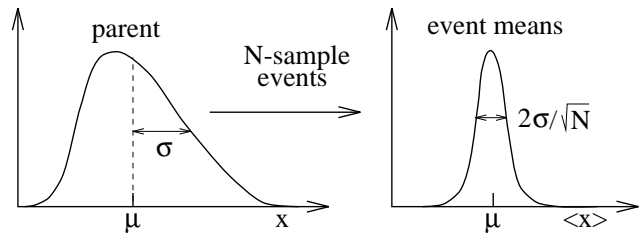}
\caption{The central limit theorem is illustrated. For an arbitrary parent distribution (left) the distribution of $N$-fold means approaches with increasing $N$ a gaussian whose width is simply related to the parent width.\label{centrala}}
\end{figure}

I begin a discussion of variance comparison measures with an elementary treatment of the central limit theorem (CLT).  In later sections we will develop a more
complete view of the CLT and variance comparisons applicable to EbyE
analysis.  The central limit theorem, first stated by deMoivre and proved
under general assumptions by Markov, plays an important role in many
statistical applications.  It is represented schematically in Fig.
\ref{centrala}.  The following is an elementary statement of the CLT as
typically encountered in a statistical context.

\begin{quote}

Given a {\em fixed}\/ arbitrary parent distribution with mean $\mu$ and 
variance $\sigma^2$, and given an ensemble of events, each composed of 
$N$-fold {\em independent}\/ samples from the fixed parent:

\begin{itemize}
\item The distribution of event means approaches a gaussian distribution 
with mean $\mu$ and variance $\sigma^2/N$. 
\item More generally, the $m^{th}$ moment about the mean of the event-mean 
distribution is reduced from the $m^{th}$ moment about the mean of the 
parent distribution by a factor $1/N^{m-1}$.
\end{itemize}
\end{quote}
The more general ($m^{th}$ moment) aspect of the theorem implies rapid asymptotic approach to a gaussian distribution (itself defined by the requirement that all moments beyond the second are zero) as $N \rightarrow \infty$.

Applied to the problem of event-wise momentum fluctuations in HI collisions as an example the CLT can be formulated in the following way. If events formed from N {\em independent}\/ p$_t$ samples with total momentum $P_t$ (an ensemble of 
final-state multiparticle distributions resulting from nominally identical 
collision events) are drawn from a {\em static}\/ parent distribution (estimated by the 
inclusive p$_t$ distribution) the distribution of event means $<\!$p$_t\!>$ = P$_t/$N approximates a gaussian distribution with mean approaching the 
parent mean $\overline{p_t}$ and variance $\sigma_{<p_t>}^2$ related to the parent variance $\sigma_{p_t}^2$ by
\begin{eqnarray} \label{clt1}
N \sigma_{<p_t>}^2  &=&  \sigma_{p_t}^2
\end{eqnarray}

Any departures from the CLT  are due to 1)
(anti)correlated samples and/or 2) variation of the parent during the
sample history.  It is case 2) that we prefer to identify with `dynamical
fluctuations,' although this is not a general convention.  The term
`dynamical fluctuations' has also been taken to represent all fluctuations
beyond finite-number statistics \cite{vol}.  Net positive correlation
implies that $N_{eff} < N$, that is, the {\em effective}\/ sample is less
than $N$.    The opposite is true for anticorrelated samples.   It is important to note that the combined presence of different correlation types and dynamical parent variations may nevertheless result in variance  behavior {\em apparently consistent} with the CLT.  This ambiguity will be explored below.

\section{The $\Phi_{p_t}$ Measure}

To facilitate the quantitative study of fluctuations and correlations in
HI collisions Ga\'{z}dzicki and Mr\'{o}wczy\'{n}ski (GM) \cite{gnm}
developed a differential correlation measure $\Phi_{p_t}$ to study
equilibration in HI collisions.  This measure and its experimental
applications have stimulated much theoretical activity in EbyE analysis
recently \cite{xxx}.  GM cite $N$-$<\!p_t\!>$ correlations in N-N
collisions \cite{kaf} as a possible means to track the equilibration
process in heavier collision systems.  For the case which prompted the
development of the measure, if A-A collision events are indeed linear
superpositions of independent N-N subevents they suggest that a suitably 
defined fluctuation measure for A-A events should have the same value as 
for N-N events taken singly if linearity prevails.

According to GM the  $\Phi_{p_t}$ width comparison measure which they define satisfies two important conditions: 1) linearity - superposing  independent and 
equivalent subsystems to form a composite system results in the same 
measure value for subsystems and composite; and 2) zero value for 
independence - if the atomic elements of the subsystems (and thus the 
composite) are themselves statistically independent the measure is zero. This measure has been proposed as a measure of EbyE $p_t$ fluctuations. It is also proposed as a general measure to detect event-wise dynamical fluctuations in energy and possibly other kinematic or chemical quantities \cite{ma2}. 

Rather than comparing variances directly $\Phi_{p_t}$ is  defined in terms of $rms$ quantities with the basic
structure 
\begin{eqnarray} \Phi_{p_t} &=& \sqrt{{N}} \sigma_{<\!p_t\!>}  -
\sigma_{p_t} 
\end{eqnarray} 
which is zero if the event ensemble is consistent with the CLT.
Following GM one can define an event-wise variation about the mean
\begin{eqnarray}
Z_e &=& \sum_{i=1}^{N_e} (x_{i} - \overline{x}) \nonumber \\
 &=&  N_e \cdot ( <\!x\!>_e - \overline{x} ) 
\end{eqnarray}
and then define the $rms$ comparison measure
\begin{eqnarray} \label{phidef}
\Phi_x &=& \sqrt{\overline{Z^2}/\overline{N}} - \sigma_x.
\end{eqnarray}
This measure satisfies both of conditions 1) and 2) above. It is zero if 
the event is composed of uncorrelated objects (elements with no 
resolved structure). The requirement 1), 2) and Eq. (\ref{phidef}) with $\Phi_x = 0$ are obviously equivalent to the central limit theorem. Given its fundamental importance it is not surprising that the CLT is frequently rediscovered. GM apply this general comparison measure to single-particle $p_t$ distributions as $\Phi_{p_t}$. To specify $\Phi_{p_t}$ one replaces $x$ by $p_t$ in the preceding.

In a $\Phi_{p_t}$ analysis of NA49 Pb-Pb data the conclusion was tentatively formed \cite{hir,gup,wuh} that ISS
and $N$-$<\!p_t\!>$ fluctuations are substantially reduced, either by collision dynamics and medium effects or by final-state rescattering, to a
point below statistical significance and certainly below values observed
in p-A and N-N systems respectively. In a further study \cite{prl} involving detailed
simulations of detector response (two-track resolution) and quantum
statistical contributions to $\Phi_{p_t}$ a model study was conducted which placed an upper limit on dynamical (parent) fluctuations at 1\% of the inclusive $p_t$ distribution $rms$ variation.

\section{The Structure of $\Phi_{p_t}$} \label{struct}

Application of the CLT in the form of $\Phi_{p_t}$ to transverse momentum
fluctuations has some unanticipated aspects.  As suggested above, there is
a number of bipolar and positive-definite contributions to $\Phi_{p_t}$.
If the measure is `zero' one does not know whether correlations are really
absent or instead somehow present in equal magnitude but opposite sign
from separate sources, or whether bipolar effects from a single mechanism
distributed over a pair-momentum space make no {\em net}\/ contribution to
this integral measure.  Even if $\Phi_{p_t}$ is nonzero these ambiguities
remain. These issues are not confronted in an elementary statement of the
CLT as presented above.  To better interpret $\Phi_{p_t}$ and to formulate
improved fluctuation measures we examine the structure of $\Phi_{p_t}$ in
detail.  An important complicating element is the presence of event-wise
multiplicity variations, whereas our simple statement of the CLT assumed a
fixed sample number $N$.  One must finally accommodate the $N$-$<pt>$
correlations which originally motivated the development of $\Phi_{p_t}$.

Examining the algebraic structure of $\Phi_{p_t}$ in more detail we  will
find that this measure is only part of a more complex scale-dependent 
covariance measure system. We will obtain a more comprehensive replacement
for the elementary CLT and a generalization of the $\Phi_{p_t}$ comparison
measure. This  fuller picture will enable us to assess possible physical
sources of  deviation from the CLT and how they are manifest in
$\Phi_{p_t}$ and  associated measures.

Because of potential correlations among statistical quantities and the need
for precision in obtaining small differential quantities from large numbers
with great statistical power I summarize the essential elementary quantities.
$\sum_e = {\cal M}$ is a sum over the event ensemble.  $\sum_i = N_e$ is a
sum over elements of event $e$.  $\sum_{e,i} / \sum_e = \overline{N}$ is
the mean event multiplicity.  $\sum_i x_i = X_e$ is an event-wise total measure.
$ \sum_i x_i/ \sum_i = X_e/N_e = <\!x\!>_e $ is an event-wise mean.
$\sum_{e,i} x_i / \sum_{e,i} = \overline{x}$ is an ensemble (inclusive)
mean over all events.  $\sum_e <\!x\!>_e / \sum_{e} =  \overline{<\!x\!>}$ 
is an ensemble-averaged event-wise mean.  In general,
$\overline{<\!x\!>} \neq \overline{x}$ because of possible correlations
between $N_e$ and $<\!x\!>_e$. In fact, 
\begin{eqnarray}
\overline{N}(\overline{<\!x\!>} - \overline{x}) = -r(<\!x\!>,N) \, 
\sigma_{<\!x\!>} \sigma_N = \sigma^2_{<\!x\!>N}
\end{eqnarray}
where the linear correlation coefficient $r(x,y)$ is defined below, and 
$\sigma^2_{<\!x\!>N}$ is a covariance.

\subsection{Variances}

This paper emphasizes global fluctuation measures based on variance. The
comparison measure $\Phi_{p_t}$ is based on an $rms$ difference.  We now
extend the elementary CLT to accomodate
event-wise multiplicity variation or variation of the sample number $N$.  We
consider the interplay among event-wise total momentum $P_e$, particle
momentum $p_i$, event-wise mean momentum $<\!p\!>_e$ and event
multiplicity $N_e$ in the spaces $G_1(P,N)$ or $G_1(P/N,N)$. To simplify the notation I omit subscript $t$ and indices $e$ and $i$ wherever confusion will not result.

A differential fluctuation measure based on variances and the CLT can be
expressed as
\begin{eqnarray}
\Phi_p &=& \sqrt{\overline{\left(P-\hat{P}\right)^2}/\overline{N}} - 
\sqrt{ \overline{\left(p - \overline{p}\right)^2}}
\end{eqnarray}
There are two possible ways to expand the variance 
$\overline{\left(P-\hat{P}\right)^2}$ if the number $N$ of samples
or particles fluctuates from event to event, depending on how one defines
the mean  value $\hat{P}$. In the first case $\hat{P} =
\overline{N}  \overline{p}$ and I denote the variance by $\sigma_P^2$. In 
the second case the mean value $\hat{P} = N_e \overline{p}$ contains 
the event-wise multiplicity. The variance $ \sigma^2_\Phi$ in the second 
case is the same $\overline{Z_e^2}$  appearing in the original definition 
of $\Phi_{p_t}$. To simplify I replace $ \overline{\left(p - 
\overline{p}\right)^2}$ with $ \sigma^2_p$ and $\overline{Z_e^2}$ with $ 
\sigma^2_\Phi$.

The variance of total momentum $\sigma_P^2$ is given by
\begin{eqnarray}
\sigma_P^2  &\equiv& \overline{(P - \overline{N}\cdot \overline{p})^2} 
\nonumber \\
&=& \overline{\sum_{i \neq j} p_i p_j} - \overline{N} (\overline{N} - 1 ) 
\overline{p}^2 + \overline{N} \sigma^2_p
\end{eqnarray}
The variance used to define the $\Phi_{p_t}$ measure is given by
\begin{eqnarray}
\sigma_{\Phi}^2 &\equiv& \overline{(P - N_e \cdot \overline{p})^2} 
\nonumber 
\\
&=&  \overline{\sum_{i \neq j} p_i p_j} - \overline{N (N - 1 )} \overline{p}^2 - 2 
\overline{p} \cdot \{\overline{N P} - \overline{N^2} \cdot 
\overline{p}\} + \overline{N} \sigma_p^2 \nonumber \\
&=& \sigma_P^2 - 2 \overline{p} \cdot \{\overline{N P} - \overline{N} \cdot 
 \overline{P}\} + \overline{p}^2 \, \sigma_N^2
\end{eqnarray}
This algebra preserves any $N$-$p_t$ correlations. No assumptions are made about the statistical interdependence of $N$ and $p_t$.

\subsection{Decomposition of $\Phi_{p_t}$} \label{decomp}

We analyze $\Phi_{p_t}$ as suggested by the structure of 
$\sigma_\Phi^2$. Recalling that $\Phi_{p_t} = 
\sqrt{\sigma_\Phi^2/\overline{N}} - \sigma_{p_t}$ we write
\begin{eqnarray}
\sigma_\Phi^2 &=& A + B + \overline{N} \sigma_p^2
\end{eqnarray}
where 
\begin{eqnarray}
A &=& \sum_{i\neq j} p_i p_j - \overline{N(N-1)} \cdot \overline{p}^2 
 \nonumber \\
 &=& \sigma_P^2 - \overline{N} \sigma_p^2 - \overline{p}^2 \, \sigma_N^2,
\end{eqnarray}
\begin{eqnarray}
B &=&  - 2 \overline{p} \cdot \{\overline{N^2 <\!p\!>} - \overline{N^2} 
\cdot \overline{<\!p\!>}\} \nonumber \\
&=& -2 \overline{p} \cdot r(<p>,N^2) \cdot \sigma_{<p>}\sigma_{N^2}  
\nonumber \\
  &=& - \sigma^2_P + \sigma_\Phi^2 + \overline{p}^2 \, \sigma_N^2
\end{eqnarray}
and express $\Phi_{p_t}$ as
\begin{eqnarray}
 2\sigma_{p_t} \cdot \overline{N} \,\Phi_{p_t}  &\approx& A + B 
\end{eqnarray}
This decomposition is nontrivial because of the unique roles played by the 
resulting $A$ and $B$ terms. $A$ is essentially a weighted integral of net 
two-point correlation , and $B$ is a measure of $N$-$<\!p_t\!>$ correlations. Each term is zero under conditions that satisfy the CLT. This decomposition can help to disentangle various physical contributions to $\Phi_{p_t}$. We now examine $A$ and $B$ in more detail.

\subsection{Two-point correlations} \label{twopnt}

We provisionally interpret term $A$ as the contribution of two-point correlations to $\Phi_{p_t}$. In the continuum limit this term can be 
written
\begin{eqnarray} \label{twopart}
A &=& \overline{\sum_{i\neq j} p_i p_j} - \overline{N(N-1)} \cdot 
\overline{p}^2  \nonumber \\
&\approx& \int p_1p_2 \cdot \{ \rho_2(p_1,p_2) - 
\rho_1(p_1) \rho_1(p_2) \} dp_1 dp_2
\end{eqnarray}
The quantity in curly brackets in the integrand is the connected two-point 
correlation function. It is zero in the absence of {\em net}\/ two-point 
correlations. The correlation function  used in HBT analysis is 
closely related. The integral is weighted by the prefactor $p_1p_2$, which 
leads to an undesirable emphasis on the large-$p$ region. Contributions to 
$A$ include, but are not limited to, quantum statistics, Coulomb effects, 
resonance decays, instrumental effects and `dynamical' fluctuations. 

\begin{figure}[th] 

\begin{tabular}{cc}

\begin{minipage}{.6\linewidth}
\epsfysize .95\textwidth
\xfig{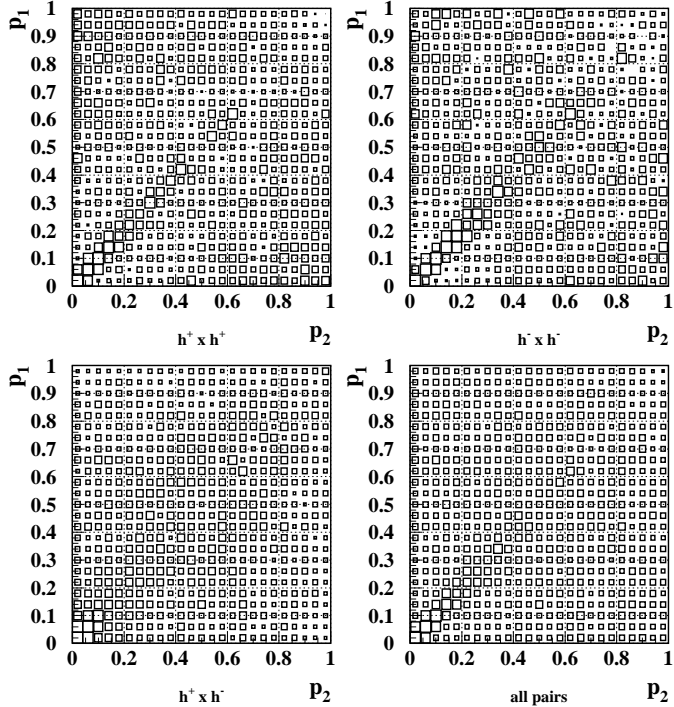}  
\end{minipage} &

\begin{minipage}{.34\linewidth}
\epsfysize 1\textwidth
\caption{Distribution of sibling to mixed pair-density ratio on $p_t 
\otimes p_t$ space for 158 GeV/nucleon Pb-Pb collisions. Density-ratio 
variation (denoted by box size) is within the interval $1 \pm 0.003$. The 
histograms are symmetric about the diagonal by construction. The pair 
momentum difference is $q$ and the momentum sum is $k$. The variables $p_i$ 
are transformed from $m_t - m_0$ in such a way as to make the statistical 
error approximately uniform over the two-point space.\label{denratio}}
\end{minipage}

\end{tabular}


\end{figure}

It is more useful to study the two-point correlations in Eq. 
(\ref{twopart}) directly rather than in terms of an integral weighted by 
the factor $p_1p_2$. One can switch to a mixed-pairs reference
\begin{eqnarray}
\rho_1(p_1) \rho_1(p_2) &\rightarrow &  \rho_{mixed}(p_1,p_2) \nonumber,
\end{eqnarray}
and compare the  two-point densities on ($p_1,p_2$) for sibling pairs 
and mixed pairs. $A$ then becomes
\begin{eqnarray}
A &\approx& \int p_1p_2 \cdot \{\rho_{sib} - \rho_{mixed}\} dp_1 dp_2
\end{eqnarray} 
We can plot the ratio of density distributions $\rho_{sib}/\rho_{mixed}$
for sibling and mixed pairs, and also project the raio onto momentum
difference $q$ and mean momentum $k$ by analogy with standard HBT
analysis. 

To minimize instrumental effects and to reveal possibly different
correlation mechanism for different pair types we plot the pair density
ratio $\rho_{sib}/\rho_{mixed}$ in Fig.  \ref{denratio} \cite{jgr}
separately for two combinations of like-sign pairs, one of unlike-sign
pairs and one for all pairs taken together.  One observes a prominent
ridge along the main diagonal ($q = 0$) for the like-signed pairs
corresponding to quantum correlations and a peak in the unlike-signed pairs
($k \approx 0$) due to the Coulomb interaction.  There is also a very
broad {\em anti}correlation at large $q$ in the like-signed pair
distributions and a broad correlation at large $q$ in the unlike-signed
pair distribution.  These last results are unexpected and illustrate the
difficulty of interpreting $\Phi_{p_t}$ in a simple manner. 
A separate treatment of positive, negative and unlike-signed
pairs is generally necessary in order to extract complete information. 
Different physics may affect each combination.

Evidence for dynamical fluctuations in the parent distribution should be common to all pair types (like
and unlike pairs),  providing an additional signature of true dynamical
fluctuations. Because $\Phi_{p_t}$ is an {\em integral}\/ correlation
measure there could be major differences between the $\rho_{sib}$ and
$\rho_{mixed}$ density distributions which nevertheless remain undetected
using $\Phi_{p_t}$.  Scale-local measures and direct analysis of individual pair spaces should provide a much more sensitive approach to the search for
dynamical fluctuations.

\subsection{$N$-$<\!p_t\!>$ correlations} 

Term $B$ is proportional to a covariance or linear correlation coefficient. \begin{eqnarray}
B &=&  - 2 \overline{p} \cdot \{\overline{N^2 <\!p\!>} - \overline{N^2} 
\cdot \overline{<\!p\!>}\} \nonumber \\
&\equiv&   - 2 \overline{p} \, r(<\!p\!>,N^2)\, \overline{N} \sigma_{<\!p\!>} \, \sigma_N \nonumber \\
 &=&  - 2 \overline{p} \, \sigma^2_{<\!p\!>N^2}
\end{eqnarray}
where the linear correlation coefficient ($lcc$) for two variables $x$ and 
$y$ is defined by
\begin{eqnarray}
r(x,y) &=& { \overline{(x - \overline{x})(y - \overline{y})} \over \sigma_x 
\cdot \sigma_y } \nonumber \\
 &=& { (\overline{x y} - \overline{x}\!\cdot\! \overline{y}) \over \sigma_x 
\cdot \sigma_y }
\end{eqnarray}
and the numerator is the covariance $\sigma^2_{xy}$.

\begin{figure}[th]
\epsfxsize .6\textwidth 
\xfig{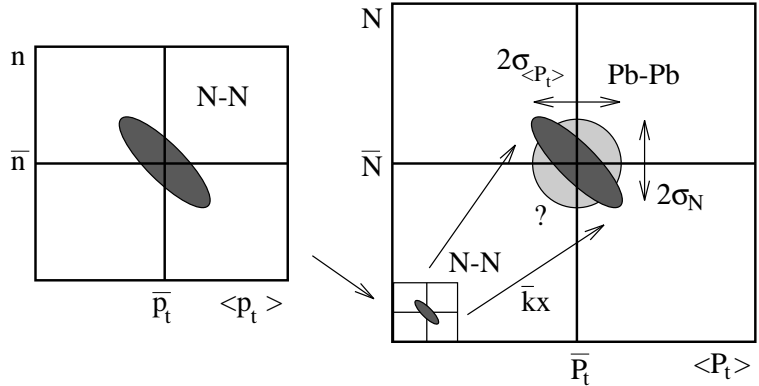}
\caption{Illustration of $N$-$<\!p_t\!>$ correlations in light systems (left) and HI collisions (right). $N$-$<\!p_t\!>$ correlations observed in light systems (N-N) may be used as a probe in HI collisions to study equilibration. The question then becomes to what extent $N$-$<\!p_t\!>$ correlations survive the equilibration process; how are conservation laws expressed differently in HI collisions than in N-N collisions? \label{lcc2}}
\end{figure}

Note that the $lcc$ also has properties 1) and 2) used in the  derivation
of $\Phi_{p_t}$. It is 1) invariant under linear composition of 
independent and equivalent subevents and 2) zero in the absence of
correlation  of event elements. To demonstrate the linearity of this
measure consider  Pb-Pb collisions as linear superpositions of N-N
collisions. If $k_e$ N-N  subevents are independently superposed in Pb-Pb
event $e$, $<\!P_t\!>_e$ is  the event-wise mean $p_t$, $<\!p_t\!>_j$ is
the mean $p_t$ for the $j^{th}$  N-N subevent and $n_j$ is its
multiplicity then:
\begin{eqnarray}
r(<\!P\!>,N)_{Pb-Pb} &=& { \overline{(<\!P_t\!> - \overline{P_t}) \cdot (N 
- 
\overline{N})} \over \sigma_{<\!P_t\!>} \cdot \sigma_{N} } \nonumber \\
 &=&  { \overline{\sum_{i,j=1}^{k} <\!p_t\!>_i\cdot n_j} - \overline{P_t} 
\cdot \overline{N} \over \sigma_{<\!P_t\!>} \cdot \sigma_{N} } \nonumber \\
 &=&  { \overline{k} \cdot \overline{<\!p_t\!>\cdot n} + \overline{k} 
(\overline{k} - 1) \cdot \overline{p_t} ~ \overline{n} - \overline{k}^2 
\cdot  \overline{p_t} ~ \overline{n} \over \overline{k} \cdot 
\sigma_{<\!p_t\!>} \cdot \sigma_{n} } \nonumber   \\
 &=& r(<\!p\!>,n)_{N-N}  \nonumber
\end{eqnarray}
Thus, the $lcc$ also offers a significant opportunity to probe 1) momentum 
conservation effects, 2) equilibration processes and 3) independence of 
particle emission.

The $lcc$ is linear under composition of partition elements (scale
invariant) because it is a ratio of total covariance matrix elements --
$\sigma^2_{xy}/\sigma_x\sigma_y$. A scale-dependent factor $M$ cancels in
this ratio. If the system is scale invariant (satisfying CLT conditions)
then this $lcc$ ratio is invariant also. However, if the CLT is
not satisfied this composite measure is not optimal for understanding
correlation sources, because it mixes the scale dependence of
variances and covariances. It is better to look at $\sigma^2_{x,y}$,
$\sigma^2_{x}$ and $\sigma^2_{y}$ or their total-covariance counterparts
separately. The intuitively motivated linearity condition
for $\Phi_{p_t}$ will become a generalized scale-invariance test of total-covariance matrix elements (cf Sec. \ref{totvar}). The question of the evolution of N-N $N$-$<$p$_t$$>$ correlations in Pb-Pb collisions which was raised in \cite{glr} and \cite{gnm} is characterized in Fig. \ref{lcc2}.

\section{A Vector Model of Variance} \label{pg}

We find that the variance comparison measure $\Phi_{p_t}$ defined in the 
spirit of the CLT apparently has two algebraic components, $A$ being the integral of net two-point correlation and $B$ containing the $lcc$ for 
$N$-$<\!p_t\!>$ correlations. In summary
\begin{eqnarray} \label{eq18}
\sigma_\Phi^2 - \overline{N} \sigma_p^2 &=& A + B \nonumber \\
 &\approx&  2 \overline{N} \sigma_p \Phi_{p_t}
\end{eqnarray}
We require a clearer picture (a geometrical model) of how the quantities $\sigma^2_P$, $\sigma^2_\Phi$ and $\sigma^2_N$ are related to one 
another and to the CLT. This problem involving measures $P$, $N$ and $P/N$ is a particular instance of the generic case  $(m_1,m_2,<m> \equiv m_1/m_2)$. 

\begin{figure}[th]
\epsfxsize .65\textwidth 
\xfig{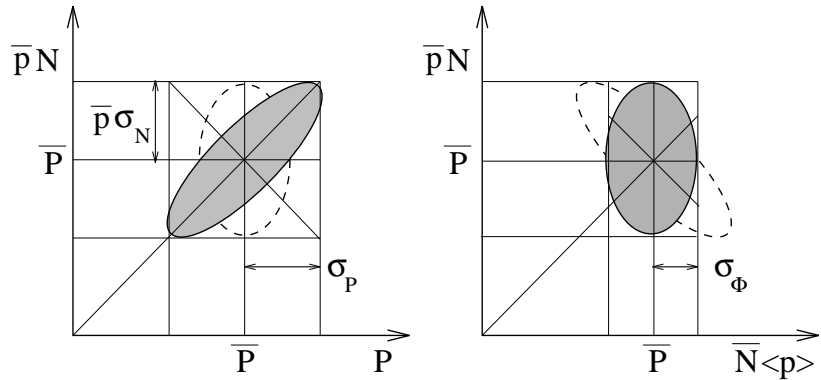}
\caption{The measure space on the left is $G_P = G_1(p,n)$ and on the right is $G_\Phi = G_1(p/n,n)$. These spaces serve as archetypes for any related set of of extensive and intensive (ratio) measures or thermodynamic state variables. The solid ellipses illustrate the case of independent sampling at small scale
followed by composition to events, whereas the dashed ellipses illustrate
independent sampling at large scale followed by partition to N objects. The
major axes of the ellipses include variations of the effective event volume, while the minor axes are determined primarily by statistical variance.\label{2d}}
\end{figure}

I now develop a  vector representation that compactly displays the algebraic relationships among the  variances. In a later section I adopt a broader description based on the covariance matrix.  To simplify notation I use a system of commensurate variables $(P,\overline{p}N,\overline{N}\!<\!p\!>)$, based on the original ($P,N,<\!p\!>$) but all having mean value $\overline{P}$ and respective $rms$ widths $(\overline{p} \sigma_N, \sigma_P, \overline{N} \sigma_{<\!p\!>} \equiv \sigma_\Phi)$. A sketch of typical 2-D density distributions on these variables with corresponding variances is shown in Fig.  \ref{2d}.  For discussion I label these single-point measure spaces $G_P$ and  $G_{\Phi}$.
The distributions on $G_P$ and  $G_{\Phi}$ are simply related algebraically. For events consisting of independent samples the covariance in $G_{\Phi}$ is small and that in $G_P$ is therefore maximal. As represented by the dashed ellipses in Fig. \ref{2d} the opposite could also be true in principle, or any intermediate case could occur. This descriptive problem underlies the $\Phi_{p_t}$-variable analysis. I now develop a vector model of variance which can be used to relate the variances and covariances in these spaces.

\subsection{A vector covariance representation} \label{npt}

\begin{figure}[th] 

\begin{tabular}{cc}

\begin{minipage}{.37\linewidth}
\epsfysize .9\textwidth
\pawplot{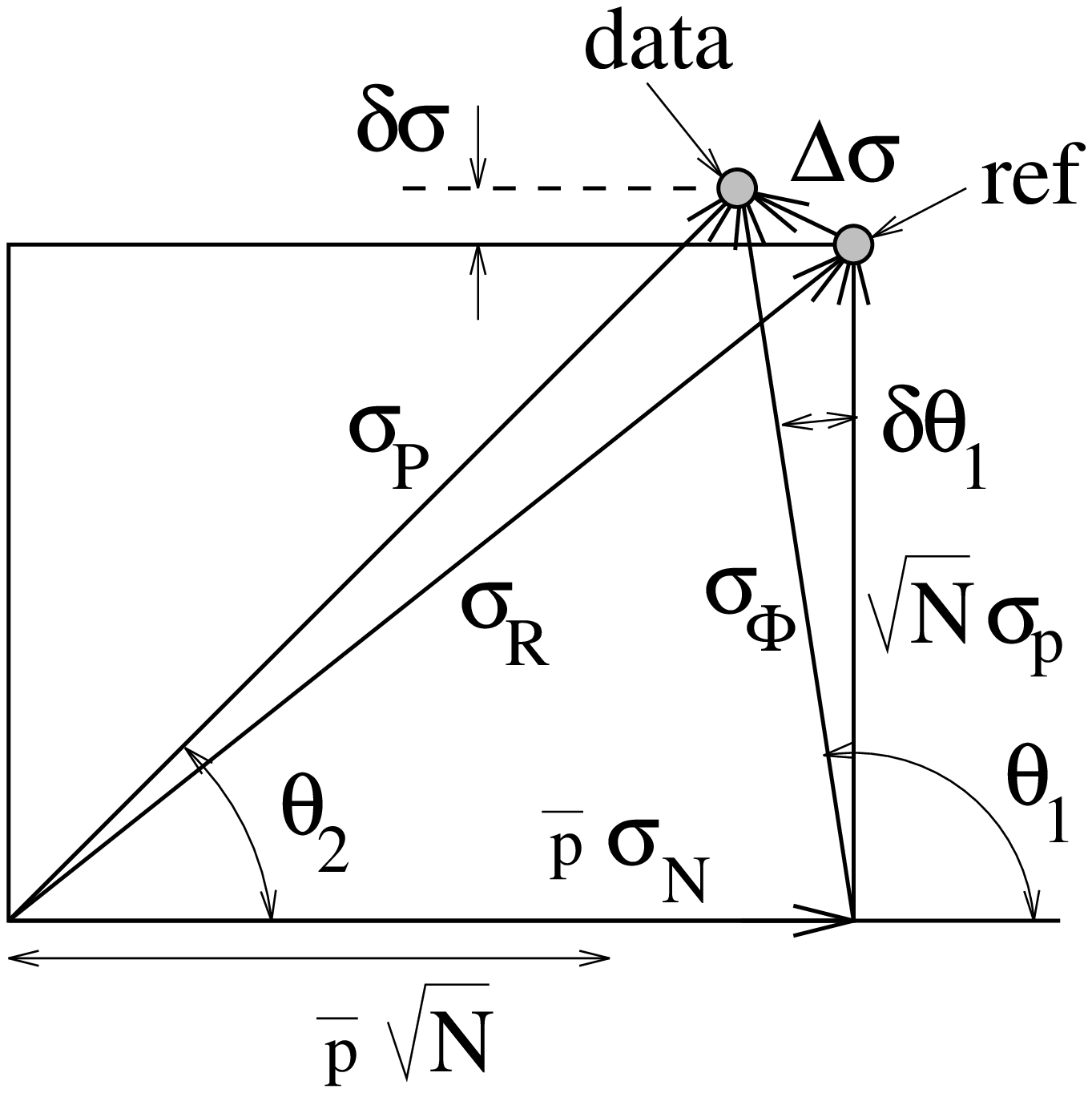}  
\caption{A vector representation relating variances in $G_P$ and $G_\Phi$ is shown above. $\sigma_R$ represents a reference, and $\sigma_P$, $\sigma_\Phi$ equivalently represent the data. The dependence of the two variance quantities on a linear correlation coefficient is shown at right.\label{sigvec}}
\end{minipage} &

\begin{minipage}{.57\linewidth}
\epsfysize 1\textwidth
\xfig{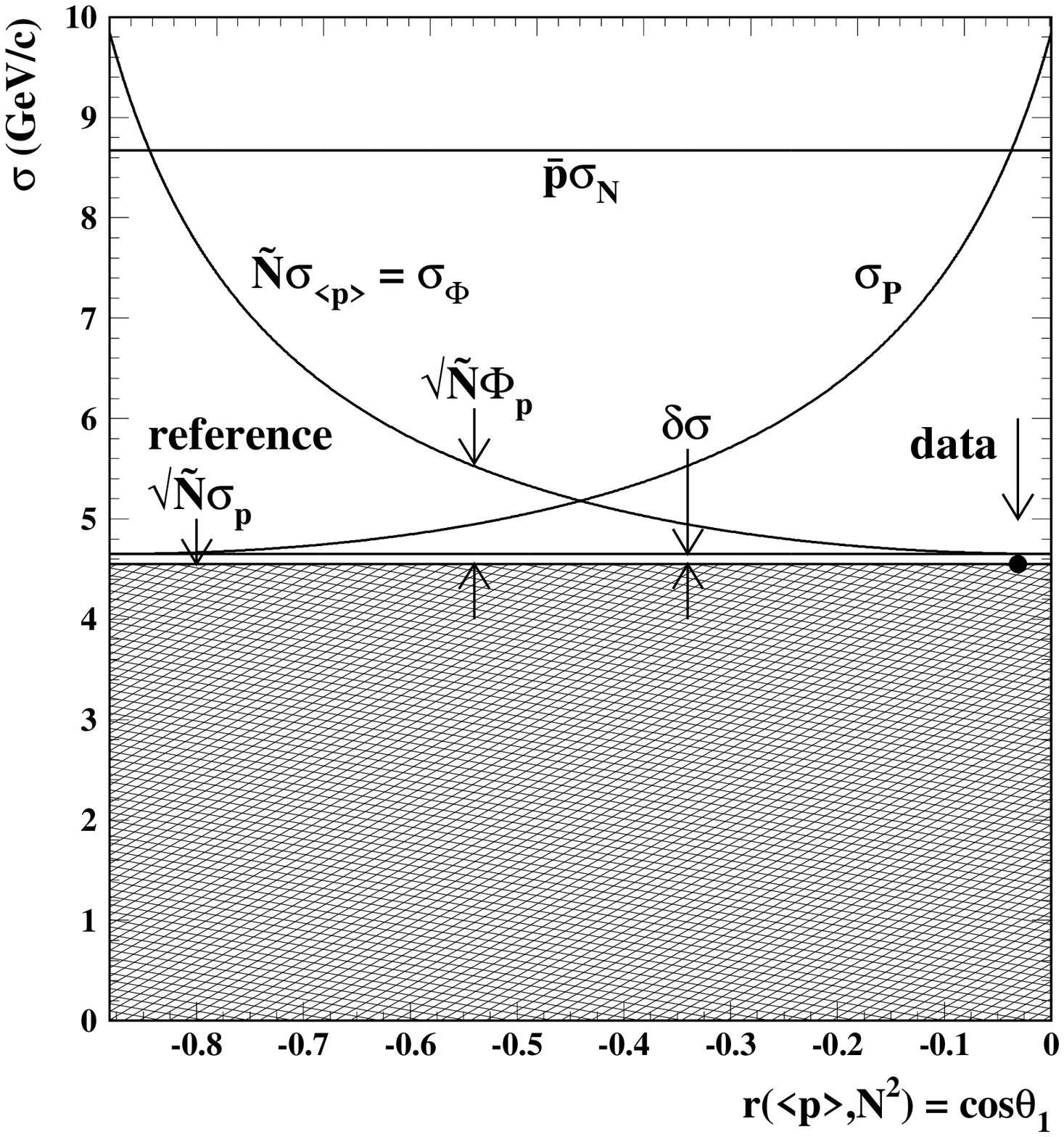}  
\end{minipage}

\end{tabular}


\end{figure}

Because of their simplicity the density distributions typically encountered in spaces $G_P$ and $G_\Phi$ for an approximately homogeneous system can be well described in terms of their first and second moments (gaussian fluctuation model), with the addition of a covariance element describing the correlation between fluctuations on two degrees of freedom. The covariance matrix in either space can be written formally as a direct product of column/row matrices whose elements are themselves vector representations of variance defined in a pseudospace. For example, we can define a generic covariance matrix
\begin{eqnarray}
{\bf K}_{xy} 
&=&
\left( 
\begin{array}[th]{ccc}
\sigma^2_x  & \sigma^2_{xy}  \\
  \sigma^2_{xy}  &  \sigma^2_y \\
\end{array}
\right)  \nonumber \\
&=&
\left( 
\begin{array}[th]{ccc}
 \vec{\sigma}_x \cdot  \vec{\sigma}_x  &  \vec{\sigma}_{x}\cdot  
\vec{\sigma}_y  \\
 \vec{ \sigma}_{x}\cdot  \vec{\sigma}_y   &   \vec{\sigma}_y \cdot  
\vec{\sigma}_y \\
\end{array}
\right)  \nonumber \\
&=&
\left( 
\begin{array}[th]{ccc}
\vec{\sigma}_{x}   \\
 \vec{ \sigma}_y  \\
\end{array}
\right) 
\cdot 
\left( 
\begin{array}[th]{ccc}
\vec{\sigma}_{x}  &  \vec{ \sigma}_y 
\end{array}
\right) 
\end{eqnarray}
where the dot product $\vec{\sigma}_{x}\cdot  \vec{\sigma}_y$ is defined by
\begin{eqnarray}
\vec{\sigma}_{x}\cdot  \vec{\sigma}_y &=& \sigma_{x} \,\sigma_y cos\theta 
\end{eqnarray}
and $cos\theta \equiv r(x,y)$ is the linear correlation coefficient between variables $x$ and $y$. 

This system can be used to relate the variances in spaces $G_P$ and 
$G_\Phi$. To do this I first form a linearized relationship between the 
two spaces. I then have 
\begin{eqnarray}
\overline{N} (P/N - \overline{p}) 
&\approx& (P - \overline{P})  - \overline{p} \, (N - \overline{N})
\end{eqnarray}
with corresponding variance vectors related by
\begin{eqnarray} \label{eq25}
 \vec{\sigma}_\Phi &\approx& \vec{\sigma}_P - \overline{p} \, 
\vec{\sigma}_N
\end{eqnarray}
(or more generally $\bar m_2 \,\vec{\sigma}_{<m>} = \vec{\sigma}_{m_1} - \bar m_1/\bar m_2 \cdot \vec{\sigma}_{m_2}$). This vector relation is illustrated on the left side of Fig. \ref{sigvec}. Examples of global intensive quantities for a hadronic system are $N_K/N_\pi$ (flavor fluctuations), $N_+/N_-$ (isovector fluctuations), $N_0/(N_+ + N_-)$ (isotensor fluctuations) and $N_{B - \bar B}/N_\pi$ (baryon-number fluctuations).

With some manipulation of Eq. (\ref{eq25}) we obtain two useful relationships
\begin{eqnarray}
\sigma_P^2 &=& \sigma_\Phi^2 + 2r_1\, \overline{p} \sigma_N \, \sigma_\Phi 
+ \overline{p}^2 \sigma^2_N
\end{eqnarray}
\begin{eqnarray} \label{sigphi}
\sigma_\Phi^2 &=& \sigma_P^2 - 2r_2\, \overline{p} \sigma_N \, \sigma_P + 
\overline{p}^2 \sigma^2_N
\end{eqnarray}
where $r_1 = r(<\!p\!>,N^2)$ and $r_2 = r(P,N)$ are the linear correlation 
coefficients. I define angles $\theta_1$ and $\theta_2$ by $r_1 \equiv cos\theta_1$ and $r_2 \equiv cos\theta_2$. The geometrical interpretation of these angles is also shown in Fig. \ref{sigvec}. By projecting out cartesian components I obtain useful sine and cosine equations
\begin{eqnarray} \label{sineqn}
\overline{p} \, \sigma_N &=&  \sigma_P\, cos\theta_2 - \sigma_\Phi \, 
cos\theta_1 
\end{eqnarray}
\begin{eqnarray} \label{eq32}
\sigma_\Phi \, sin\theta_1 = \sigma_P \, sin\theta_2 = \sqrt{N} 
\, \sigma_p + \delta \sigma
\end{eqnarray}
the second of which includes a possible contribution $\delta \sigma$ from an unspecified variance source. Eqs. (\ref{sigphi}) and (\ref{sineqn}) are analogs to Eqs. (17) and (23)  respectively of \cite{edw}. In the present case the results assume no  special conditions and arbitrary deviation from the CLT whereas the results in \cite{edw} are derived in the specific context of an equilibrated  pion gas.

\subsection{Vector deviation from a CLT reference}

The vector description of variance in spaces $G_P$ and $G_\Phi$ has so far
been generic. That is, $-cos\theta_1$ and $cos\theta_2$ could have taken
any consistent pair of values on the interval $[0,1]$. The two spaces are
complementary and interchangeable. A reference system for the CLT applicable to HI collisions can now
be defined by $\vec{\sigma}_R =  [\overline{p}\sigma_N,\sqrt{N}\sigma_p]$.
This choice implies zero covariance between variables $<\!p_t\!>$ and $N$ ($cos\theta_1 = 0$) and momentum
variance given by a CLT expectation The number variance $\sigma^2_N$ exceeds the Poisson expectation $\bar N$ primarily due to correlations from resonance decays.  This reference is consistent with
the central limit theorem but represents a more complete
description (including covariance). The physical interpretation of this reference is consistent with the CLT -- individual particle momenta chosen independently from a fixed parent distribution represented by an inclusive $p_t$ distribution.

Given this reference choice I can define a vector measure of deviation from the CLT reference, vector $\Delta \vec{\sigma}$ running  from the reference ($\vec{\sigma}_R$) to a point  representing the data ($\vec{\sigma}_P$)
\begin{eqnarray} \label{thing}
\Delta \vec{\sigma} &\equiv& \vec{\sigma}_P - \vec{\sigma}_R \nonumber \\
&=& [\sigma_\Phi cos\theta_1,\sigma_\Phi sin\theta_1-\sqrt{N}\sigma_p]\nonumber \\
&\approx& \sqrt{N} [\sigma_p \, \delta \theta_1, \Phi_{p_t} - \sigma_p \, {\delta \theta^2_1 / 2}] \nonumber \\
&\approx& \sqrt{N} [\sigma_p \, \delta \theta_1, \delta \sigma]
\end{eqnarray}
where we finally retain only terms linear in small quantities ($\delta 
\theta_1,\delta \sigma$).
We see that {\em two}\/ quantities, $\delta \theta_1 \approx cos\theta_1 = lcc  \propto B$ and $ 
\delta \sigma \propto \Phi_{p_t}$, which both have the properties 1) -- 
linearity and 2) -- zero for independent objects, provide a measure of 
deviation from a dynamical variance reference relevant to the CLT. This 
extends the initial concept of the $\Phi_{p_t}$ measure finally to handle also $N$-$<\!p_t\!>$ correlations.

Dependence of the variances on $r_1 = cos\theta_1$ is shown on the right side of Fig.
\ref{sigvec}. Note that $\Phi_{p_t}$ depends {\em quadratically}\/ on
$r(<\!p\!>,N^2) = cos \theta_1$, which is just the measure sensitive to $N$-$<\!p\!>$ correlations. This dependence was first suggested by numerical simulations \cite{gu1} and implies that $\Phi_{p_t}$ itself is ironically quite insensitive to $N$-$<\!p\!>$ correlations. 
Combining Eqs. (\ref{eq18}), (\ref{eq25}) and (\ref{eq32}) I have
\begin{eqnarray}
\Phi_{p_t} &\equiv& \sigma_\Phi / \sqrt{N} -  \sigma_p \nonumber \\
&\approx& {\sigma_p \over 2} cos^2\theta_1 + \delta \sigma.
\end{eqnarray}
which is consistent with Eq. (\ref{thing}).
We conclude that for nearly symmetric systems ($ \vec{\sigma}_P \approx \vec{\sigma}_R $) $\Phi_{p_t}  $ is primarily sensitive to {\em variance}\/ sources ($\delta \sigma$), whereas {\em covariance}\/ sources such as $N$-$<\!p\!>$ correlations must be measured explicitly by a covariance or linear correlation coefficient. In Sec. \ref{covmat} we will consider the relationship between covariance matrices in $G_P$ and $G_\Phi$ and similar spaces more generally.

A concern is raised in \cite{wuh} that $ \sigma^2_{<\!p_t\!>} \equiv \sigma^2_{P_t/N} \neq \sigma^2_{\Phi} / \overline{N}^2$, that is $\Phi_{p_t}$ is not directly related to $ \sigma^2_{<\!p_t\!>}$. But it is easy to show that the latter quantities in the present case are arbitrarily close, to the extent that the relationship between spaces $G_P$ and $G_\Phi$ can be linearized. This depends in turn on the degree to which the CLT conditions are met and on the size of the total multiplicity $\overline{N} $ (equivalently the scale interval from particle to event).

\subsection{Significance of $A$, $B$ and $\Phi_{p_t}$}

 We now reexamine $A$, $B$ and $\Phi_{p_t}$ in light of this more complete
vector variance representation. The relationships among $A$, $B$ and
$\Phi_{p_t}$ were first suggested by  numerical simulations \cite{gu1}
which  demonstrated a strong correlation between $A$ and $B$ when
$N$-$<\!p_t\!>$  correlations were artificially introduced into a
mixed-event reference. $A$ and $-B$ varied together over a $40 \, MeV/c$
range while $\Phi_{p_t}$ fluctuated about zero.  This suggested either
that $A$ and $B$ were redundant or that one carried a subset of the
information contained in the other. The relationship $A + B  \approx 2
\overline{N} \sigma_p \Phi$ simply defines a plane in a 3-D space of
$(A,B,\Phi)$ without giving emphasis to any one of the three related 
quantities. We would like to develop a more substantial understanding of
this  relationship. 

Given the definitions of $A$, $B$ and $\Phi$ I can express $A$ as
\begin{eqnarray}
A &\approx&  \sigma_\Phi^2 - \overline{N} \sigma_p^2 + 2 \, \overline{p} 
\sigma_N 
\, \sqrt{N} \sigma_p \, cos\theta_1 \nonumber \\
&\approx&  2 \sqrt{N} \sigma_p \cdot \sqrt{N} \Phi  +    2 \overline{p} 
\sigma_N  \cdot \sqrt{N} \sigma_p cos\theta_1 \nonumber \\
 &\approx& 2  \Delta \vec{\sigma} \cdot \vec{\sigma}_R
\end{eqnarray}
$A$ thus has the form of a vector dot product. Note that if the deviation 
$\Delta \vec{\sigma}$ is orthogonal to the reference $\vec{\sigma}_R$
there is no {\em  net}\/ two-point correlation, even though the CLT is
not satisfied. We now have a relationship between departures from the CLT
on the one hand  ($cos \theta_1$ and $\Phi_{p_t}$), which are parts of a global measure system, and net  two-point correlations on the other ($A$), which refer to the  microscopic particle complement. This vector result {\em suggests}\/ that $A$ is in  a measure class distinct from $cos \theta_1$ and $\Phi_{p_t}$. We will see below that this is indeed the case. In pursuing the structure of $\Phi_{p_t}$ we have discovered a  broader organization.

We also find that the presence of two-point correlations {\em may}\/
result in deviation from a reference variance system, but not
necessarily.   On the other hand deviation from a reference system {\em
may}\/ result in net two-point correlation, but again there is no
necessary relationship. Variance changes and $N$-$<\!p_t\!>$ correlations (covariances) are global effects with possible thermodynamical interpretations. Both {\em may}\/ correspond to net two-point correlations at the microscopic level or {\em  vice versa}, but there is no necessary connection. This picture is still  not fundamental. We require a more complete treatment of variance and correlation, including explicit scale dependence, to fully interpret any  results. 

This treatment of variance provides a geometrical model relating archetypal spaces $G_P$ and $G_\Phi$. It illustrates how small incremental contributions to variance and covariance relative to a reference are related to the $\Phi_{p_t}$ measure. It illustrates dramatically the reason that $\Phi_{p_t}$ is insensitive to $N-<\!p_t\!>$ correlations, which are instead measured by a covariance element. We have established that the decomposition of $\Phi_{p_t}$ into components $A + B$ is nontrivial, that in fact $\Phi_{p_t}$ itself approximates a component of a more comprehensive measure system, and that its value is {\em not}\/ related in a necessary way to the presence or absence of system correlations, either differential or integral. To proceed further we must
extend our description to a system of scale-dependent covariance matrices.

\section{Scale-Dependent Variance} 

To address the ambiguities encountered in the $\Phi_{p_t}$ analysis we now attempt to extract more complete information from collision
events. We do this by expanding the description in two ways: 1) we
move to an explicit scale-dependent description, and 2) we adopt a
description based on the complete covariance matrix for all additive
measures relevant to the dynamical system. The vector representation of variance developed previously is contained in a larger description involving
scale-dependent covariance matrices and correlation integrals.

There are four spaces involved in a variance study. $P_1(\vec{x})$ is the single-point primary space with measure-density system $\vec{m}$ defined on it. $Q_2(\vec{x} \otimes \vec{x})$ is the corresponding two-point space with measure product density $\vec{m} \otimes \vec{m}$ defined on it. These spaces may not be directly accessible to observation. Measure {\em integrals} for specific partition systems are themselves distributed on spaces $G_1(\vec{m})$ and $G_2(\vec{m} \otimes \vec{m})$ which may represent the results of direct measurement. The covariance matrices described below represent a gaussian approximation to distributions on $G_1(\vec{m})$. Covariance-matrix differences are in turn related to correlation integrals $\Delta C_2(m_1,m_2;\delta x_1,\delta x_2)$ over two-point space $Q_2(\vec{x} \otimes \vec{x})$ and covariance integrals $A(\delta x_1, \delta x_2; m_1, m_2)$ over space $G_2(\vec{m} \otimes \vec{m})$.

\subsection{Variance and correlation integrals}

A variance is implicitly defined at a particular scale (the bin size of a
binning system).  Each entry in Fig.  \ref{2d} corresponds to an event,
which can also be considered as a partition element or `bin' in some
larger space (Sec. \ref{ensemb}) resulting from the sampling of a general
thermodynamic system at some arbitrary scale.  We now move to a more
explicit and general scale-dependent binning approach, with events and
particles considered as particular cases.

I assume a space $\Omega$ in which particles, events and event ensembles
are arrayed. $\Omega$ is the primary or $P$ space. This space is binned
at some arbitrary scale $\delta x$.  At some smaller scale $a$ bins may be identified with
particles.  At a larger scale $L$ bins may coincide with events.  The
space is arbitrarily large and may contain many events (an event
ensemble). Associated with any $P$ space is a set of $Q_q$ spaces consisting of $q$-fold cartesian products of $P$ space. In the discussion of variances that follows the space $Q_2$ will be of primary importance. This space represents all the point pairs of $P$ space.

I consider relationships among measures on a bounded scale interval
$[\delta x_1,\delta x_2]$ on which I define a scale-dependent covariance
matrix ${\bf K}(\vec{m},\delta x)$. ($\vec{m}$ is the set of measures defined on $\Omega$).  This scale-local generalization with particles
and  events as limiting cases provides a more complete and
self-consistent  description of variance.  I begin with the scale
dependence of variance for a single measure.

Consider an additive measure (say momentum) distributed on $\Omega$ and 
binned at two scales $(\delta x_1,\delta x_2)$, with typical bin contents 
$p(\delta x_1)$ and $p(\delta x_2)$ and with $p(\delta x_2) = \sum_{i=1}^N 
p_i(\delta x_1)$ for the smaller bins contained in a larger bin. I can
then write for each large-scale bin
\begin{eqnarray}
p^2(\delta x_2) &=& \sum_{i=1}^{N} p_i^2(\delta x_1) + \sum_{i \neq j = 
1}^{N} p_i(\delta x_1) p_j(\delta x_1)
\end{eqnarray}
If I average over all bins I have
\begin{eqnarray}
\overline{p^2}(\delta x_2) -  {N} \, \overline{p^2}(\delta x_1) &=&  
\overline{\sum_{i \neq j} p_i(\delta x_1) p_j(\delta x_1)}
\end{eqnarray}
where ${N}(\delta x_1 ,\delta x_2) = (\delta x_2 / \delta x_1)^d =
M(\delta  x_1)/M(\delta x_2)$ is the number of smaller bins
contained in a  larger bin ($d$ is the P-space dimension, which I hereafter omit to simplify notation). The RHS is also expressible as the integral over pair-momentum space  of a two-point density distribution. I discuss this integral in more  detail below.  The LHS can be expressed as the  difference between two rank-2 correlation integrals.

The rank-$q$ normalized correlation integral at scale $\delta x$ for an 
additive measure $p$ can be approximated by
\begin{eqnarray} 
C_q(\delta x) &\approx &  \sum_{i=1}^{M(\delta x)} 
p^q_i(\delta x) / \left\{\sum_{i=1}^{M(\delta x)} 
p_i(\delta x)\right\}^q
\end{eqnarray}
where  $M(\delta x)$ is the number of {\em occupied}\/ bins at scale
$\delta  x$ in the space $\Omega$ (containing nonzero measure). This
bin-based definition approximates an integral on scale in space $Q_q$ from
zero scale up to the bin size $\delta x$. For an uncorrelated or uniform
distribution $C_q(\delta x) \rightarrow C_{q,ref} =  M^{1-q}(\delta x)$,
which is one possible reference choice. 

\begin{figure}[th]
\vskip .2in
\begin{tabular}{cc}

\begin{minipage}{.57\linewidth}
\epsfxsize .7\textwidth
\pawplot{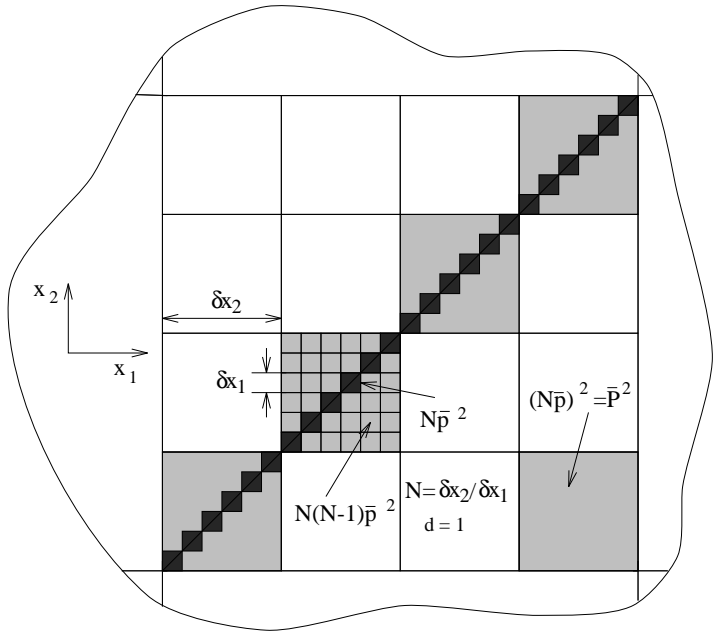}  
\end{minipage} &

\begin{minipage}{.37\linewidth}
\epsfysize 1\textwidth
\caption{Two-point space $Q_2(\vec{x} \otimes \vec{x})$ corresponding to a 1-D primary space. Statistical quantities are given for a uniform reference distribution. $Q_2$ is binned at three arbitrary scales. A comparison is made in the text between the variance contained in bins at a smaller scale (dark) and at a larger scale (light).\label{grid2}}\end{minipage}

\end{tabular}

\vskip .3in

\end{figure}

I can express  mean-square quantities in terms of the correlation integral by
\begin{eqnarray}
\overline{p^2}(\delta x) &\approx& M(\delta x) \overline{p}^2(\delta x) 
C_2(\delta x)
\end{eqnarray}
from which I obtain
\begin{eqnarray} \label{blah}
 \overline{p^2}(\delta x_2) - N \overline{p^2}(\delta x_1) 
 &\approx& M(\delta x_2) \, \overline{p}^2(\delta x_2) C_2(\delta x_2) - 
[M^2(\delta x_1)/M(\delta x_2)]\, \overline{p}^2(\delta x_1) C_2(\delta 
x_1) 
\nonumber \\
 &\approx& \overline{p}^2(\delta x_2) \, M(\delta x_2) \{C_2(\delta 
x_2) - C_2(\delta x_1)\}\nonumber \\
 &\approx& \overline{p}^2(\delta x_2) \, M(\delta x_2) \, C_2(\delta 
x_1,\delta x_2)
\end{eqnarray}
The difference between correlation integrals at two scales is just a
correlation  integral over a scale interval, $C_2(\delta x_1,\delta x_2)$.
This  justifies the observation made earlier that at least some global
variables, those based on moments,  are scale integrals. If we subtract
from both sides of Eq. (\ref{blah}) the  corresponding expressions for a
reference  distribution such as a uniform distribution, with
$\overline{p^2} =  \overline{p}^2$ and $\overline{\sum_{i \neq j} p_i p_j}
= N(N-1)  \overline{p}^2$ as shown in Fig. \ref{grid2}, then we have the
variance difference
\begin{eqnarray} \label{blah2}
\sigma^2(p,p,\delta x_2) - N \sigma^2(p,p,\delta x_1) 
 &=& \overline{p}^2(\delta x_2) M(\delta x_2) \, \Delta C_2(p,p,\delta 
x_1,\delta x_2) 
\end{eqnarray}
where we adopt the notation $(m_1,m_2,\delta x)$ to anticipate generalization
to covariance matrices for arbitrary measure systems. $\Delta C_2(p,p;\delta x_1,\delta x_2)$ is then the difference between 
correlation integrals over a bounded scale interval for object and
reference distributions. The implicit reference for a standard variance is a uniform distribution on the system of occupied bins or {\em support}\/ of the object distribution. The variance quantities $\sigma^2(p,p,\delta x)$ could also be defined in terms of an arbitrary nonuniform reference. 

Fig. \ref{grid2} illustrates several of these concepts. Bins at smaller (dark) and larger (light) scale are highlighted along the main diagonal of $Q_2(\vec{x} \otimes \vec{x})$  space. The variation of {\em total variance} (next section) with scale implicit in Eq. (\ref{blah2}) can be illustrated with the following argument.  $\sigma^2(p,p,\delta x_1)$ represents the variance per bin for the smaller bins, while $\sigma^2(p,p,\delta x_2)$ represents the same quantity for the larger bins. Since there are $N$ smaller bins in a larger bin $\sigma^2(p,p,\delta x_2) - N \sigma^2(p,p,\delta x_1)$ represents the {\em excess} variance in a larger bin not accounted for by that contained in the smaller bins. This excess must be contributed by the $N(N-1)$ off-diagonal smaller bins contained in a larger bin. Within a factor, the measure of additional variance in the off-diagonal bins is the correlation integral difference $\Delta C_2(p,p;\delta x_1,\delta x_2)$ over the scale interval bounded by the two bin sizes. The two diagonal arrays of bins {approximate} the integration regions (diagonal strips) appropriate for exact correlation integrals $C_2(\delta x_1)$ and $C_2(\delta x_2)$ in $Q_2$. This connection of variance with correlation integrals is in part motivated by subsequent integration of covariance treatments with scale-local topological measures \cite{re1,re2,sca}.

I now return to the cross term
\begin{eqnarray}
\overline{\sum_{i \neq j} p_i p_j} 
&\approx& \int p_1p_2 \cdot  \rho_2(p_1,p_2) \, dp_1 dp_2
\end{eqnarray}
This relation connects the two-point measure space $G_2(p,p)$ (right side)
with the two-point space $Q_2(x,x)$ represented in Fig. \ref{grid2} (left
side). The integral over the entire two-point momentum space of the
density $\rho_2(\delta x_1,\delta x_2;p_1,p_2)$ is equal to a correlation
integral over a {\em specific scale interval}\/ in $Q_2$ of the density
$\rho_2(p,p;x_1,x_2)$. Distribution $\rho_2(\delta x_1,\delta
x_2;p_1,p_2)$ is a parametric function of that scale interval as indicated by the argument. Note the convention $\rho_2(params;variables)$ where
$params$ determine the nature of the density and $variables$ are the
position indices for the corresponding space.

It is useful to adopt the convention that symbol $a$ represents a  `particle' scale and $L$ represents an `event' scale, with $\Omega$ representing the entire containing space. In the particle-event limit $N(\delta x_1,\delta x_2) \rightarrow \bar N_{part}$, and $M(\delta x_2) \rightarrow {\cal M}$ (total event number) as $(\delta x_1,\delta x_2) \rightarrow (a,L)$. The distribution in $G_P = G_1(p,n)$ is then $\rho_2(L,\Omega;p,n)$, and the two-point particle momentum distribution in $G_2(p_1,p_2)$ is $\rho_2(a,L;p_1,p_2)$.  We can generally write $\rho_2(\delta x_1,\delta x_2;m_i,m_j)$ for the density in $G_2(m_1,m_2)$, consistent with the notation above. To  complete the treatment we write 
\begin{eqnarray} \label{adef}
\sigma^2(p,p,\delta x_2) - N \sigma^2(p,p,\delta x_1)  &=& \left\{
\overline{\sum_{i \neq j} p_i p_j} \right\}_{obj} - \left\{ 
\overline{\sum_{i \neq j} p_i p_j}\right\}_{ref} \nonumber \\ 
&\approx& \int p_1p_2 \cdot \{ \rho_{2,obj}(\delta x_1,\delta x_2;) - 
\rho_{2,ref}(\delta x_1,\delta x_2;) \} dp_1 dp_2 \nonumber \\ 
&\equiv& A(\delta x_1,\delta x_2;p,p) 
\end{eqnarray}
which defines $A(\delta x_1,\delta x_2;p,p)$ as an integral over the two-point measure space $G_2(p_1,p_2)$, the same $A$ which appears in the decomposition of $\Phi_{p_t}$. In case $\rho_{2,ref}(\delta x_1,\delta x_2;)$ corresponds to a uniform reference then $A(\delta x_1,\delta x_2;p,p)$ is the {\em covariance} of the two-point density $\rho_{2,obj}(\delta x_1,\delta x_2;p_1,p_2)$.
            
This result for variance $\sigma^2(p,p,\delta
x)$ applies  also to other elements of any covariance matrix for a
measure system  $\vec{m}$. Thus, the covariance difference
$\sigma^2(p,n,\delta x_2) - N  \sigma^2(p,n,\delta x_1)$ equals an
integral over the space $G_2(p,n)$ of a  corresponding two-point density
difference $\rho_{2,obj}(\delta x_1,\delta x_2;p,n) -  
\rho_{2,ref}(\delta x_1,\delta x_2;p,n)$ represented by $A(\delta x_1,\delta x_2;p,n)$.  
       
\subsection{Total variance} \label{totvar}

The linearity property invoked in the derivation of $\Phi{p_t}$ is actually an implicit invocation of {\em scale invariance}. The approximately invariant measure is the total variance introduced in the previous section. Recalling that $N(\delta x_1,\delta x_2) = M(\delta x_1)/M(\delta x_2)$ and multiplying through Eq. (\ref{blah2}) by $M(\delta x_2)$ (omiting 
subscripts) I have
\begin{eqnarray} \label{sig}
\Sigma^2(p,p,\delta x_2)  - \Sigma^2(p,p,\delta x_1)
&\approx&P^2 \, \Delta C_2(p,p;\delta x_1,\delta x_2)
\end{eqnarray}
where $\Sigma^2(p,p,\delta x) \equiv M(\delta x) \,\sigma^2(p,p,\delta
x)$  is the {\em total variance}\/ of a measure $p$ on $\Omega$ relative to
a uniform  reference at scale $\delta x$ ($\sigma^2(p,p,\delta x)$ is the
{\em per-bin}\/ variance), and $P \equiv M(\delta x) \, \overline{p}(\delta
x)$  is a scale-invariant {\em total measure}\/ on $\Omega$. 
We establish here the concept that variation of total (co)variance on scale is the underlying issue for the central limit theorem and any correlation analysis based on global variables. Variation of the scale-dependence of total variance may be due to changes in the underlying system Lagrangian for certain constraint configurations (the neighborhood of a phase boundary), in which a correlation onset may appear or disappear, producing large fluctuations in the primary $DoF$ which determine the basic statistical fluctuations for all observables.

\begin{figure}[th]
\vskip .2in
\begin{tabular}{cc}

\begin{minipage}{.57\linewidth}
\epsfxsize .8\textwidth
\pawplot{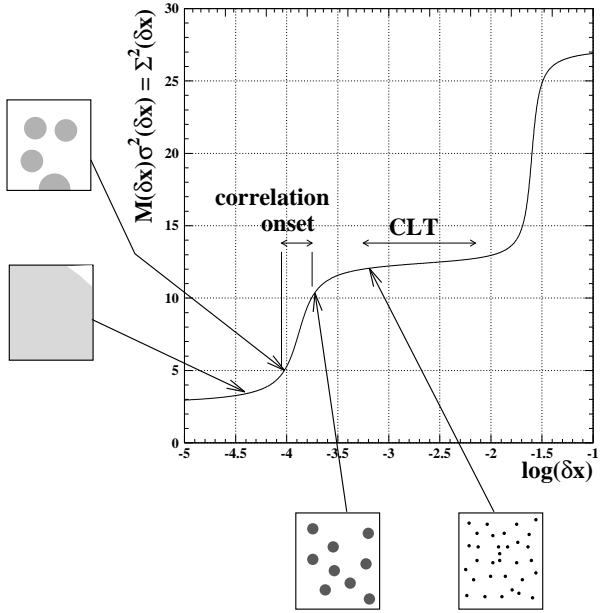}  
\end{minipage} &

\begin{minipage}{.37\linewidth}
\epsfysize 1\textwidth
\caption{Scale-dependence of total variance for a system with scale-localized correlation onsets. Correlation onset extends from the characteristic size of an object to the mean spacing between objects where localized objects predominate. The distribution of total variance is not necessarily monotone increasing with scale due to the possibility of anticorrelation in distributions. Such curves may vary with global constraints such as $(T,\mu)$ near a phase boundary, leading to departure from a CLT expectation. \label{central1a}}
\end{minipage}

\end{tabular}

\vskip .3in

\end{figure}

We can connect increased total variance with its source as represented by the correlation integral by developing a differential equation. We simply take Eq. (\ref{sig}) to the limit of small scale interval
\begin{eqnarray} \label{deq}
\partial \Sigma^2(p,p,\delta x)/ \partial \delta x
&\approx& P^2 \, \Delta \partial C_2(p,p;\delta x)/\partial \delta x
\end{eqnarray}
where $\partial C_2(p,p;\delta x)/\partial \delta x $ is an autocorrelation density. This relation then implies that any variation in total variance over a scale interval (the object of a CLT test) corresponds to a nonzero difference between object and reference autocorrelation densities over that interval. Since the autocorrelation density is a quantitative symmetry measure we see that scale-invariance of total variance is intimately related to the scale-dependent {\em relative symmetry} of object and reference distributions.

The CLT provides a useful reference for measure systems in which localized regions of correlation onset are separated by significant scale intervals with negligible accumulation of total variance (approximate scale invariance) as illustrated in Fig. \ref{central1a}. Distinguishable objects that result from a localized correlation onset can be designated `particles' or dynamical $DoF$, and correlation analysis based on integral quantities (global variables) is then simplified to charcterizing any (relatively small) total variance accumulated over some scale interval between these correlation onsets.

Approximate invariance of total variance over a bounded scale interval is espressed by $\Sigma^2(\delta x_2) \approx \Sigma^2(\delta x_1)$ for some $[\delta x_1, \delta x_2]$. We then have
\begin{eqnarray}
N(\delta x_1,\delta x_2) \, \sigma^2_{<\!p_t\!>} 
&\equiv&  \sigma^2_{P_t}/N(\delta x_1,\delta x_2) \nonumber \\
&=& \Sigma^2(\delta x_2)/M(\delta x_1)  \nonumber \\
&\approx&  \Sigma^2(\delta x_1)/M(\delta x_1) \nonumber \\
&\approx& \sigma^2_{p_t}
\end{eqnarray}
which is a statement of the CLT as used in the definition of the $\Phi_{p_t}$ variable (Eq. (\ref{clt1})).

In summary, the {\em accumulation}\/ of total variance on scale is a basic issue for any global-variables analysis based on a CLT test. This is simplest to study in cases where one has scale-localized correlation onsets separated by scale intervals of nearly constant total variance, over which a CLT reference is a good approximation. We have shown - Eq. (\ref{adef}) -- that the difference $\sigma_P - N \, \sigma_p$ is actually a scale integral represented by $A(\delta x_1,\delta x_2;p,p)$, and that it is desirable to examine the integrand of $A$ directly in a differential analysis whenever possible. An extention of this program to {\em scale-local}\/ topological measures examines the detailed scale-depence of total-variance accumulation or its equivalent without prior assumptions \cite{chi}. With a scale-dependent treatment of variance in hand we now proceed to a general system of scale-dependent covariance matrices.

\section{Scale-Dependent Covariance Matrices} \label{covmat}

We now consider an arbitrary system of dynamical measures and the corresponding scale-dependent covariance matrix. A covariance matrix provides a gaussian approximation to a scale-dependent density distribution in measure space $G_1(\vec{m})$. As we have shown, variation of (co)variances on scale can be equated to corresponding integrals over two-point measure spaces. We distinguish between correlation integrals of densities $\rho_2(m_1,m_2;x_1,x_2)$ on spaces $Q_2(x_1,x_2)$ represented by $C_2(m_1,m_2;\delta x_1,\delta x_2)$ and integrals of densities $\rho_2(\delta x_1,\delta x_2;m_1,m_2)$ on two-point measure spaces $G_2(m_1,m_2)$ represented by $A(\delta x_1,\delta x_2;m_1,m_2)$.

\subsection{Single-point and two-point measure spaces}

If a space $P_1(\vec{x})$ supports a set of additive measures $\vec{m} = (m_1,m_2,...)$ ({\em e.g.,} momentum $p$ and multiplicity $n$) having bin values $\vec{m}_i(\delta x)$ for the $i{th}$ bin of size $\delta x$ represented as a density in a single-point measure space $G_1(\vec {m})$ with projections onto pairs of measure values such as $G_1(p,n)$, and I form a two-point space $Q_2(\vec{x} \otimes \vec{x})$ (Fig.  \ref{grid2}) then I can define a corresponding general two-point measure space $G_2(\vec{m}_1 \otimes \vec{m}_2)$ with projections onto pair spaces such as $G_2(p_1,p_2)$, $G_2(p_1,n_2)$, $G_2(n_1,n_2)$ and so forth.  Each bin in $Q_2$ or pair of bins in P space contributes one `point' (integral of the measure densities over the bin) to a density distribution $\rho_2(\delta x_1,\delta x_2;m_1,m_2)$ {\em for a specific scale interval}\/ in each of the two-point measure spaces (these spaces are binned also, at the resolution of a detection device).  These density distributions are parameterized by the scale intervals in P space used to define them.

Specifying a scale {\em interval}\/ ($\delta x_1,\delta x_2$) may seem 
mysterious. The lower limit is obviously the bin size in P space used to 
define the densities. But to form a correlation space one must specify
over  what region of P space bin contents will be correlated (Fig. 
\ref{grid2}). The size of this  region (a type of correlation length)
provides the upper limit to the scale interval. In an EbyE analysis
`particles' are correlated within `events' and not between events over an 
entire ensemble. 

Measures defined at the event scale may be used to sort events into categories. One aspect of EbyE analysis is to establish such categories. Subsequent EbyE analysis may indicate that a measure space is incomplete by revealing previously unknown correlation structures within events. In defining references for correlation analysis it is desirable to form reference pairs from events that are `nearest neighbors' in this expanded measure space, for instance, events with the closest total multiplicities or transverse energies.

The quantity $A(\delta x_1,\delta x_2;p,p)$ defined in Eq. (\ref{adef}) is
an example of an integral over a two-point measure space. It is just the
term $A$ identified in  the decomposition of the $\Phi_{p_t}$ measure. It
is the integral over  measure space $G_2(p_1,p_2)$ of the connected two-point
correlation  function $\rho_{2,obj}(a,L;p_1,p_2) -
\rho_{2,ref}(a,L;p_1,p_2)$ (Fig. \ref{denratio}). Similar integrals for
various additive measures taken pair-wise form the elements of the integal
matrix ${\bf A}(\delta  x_1,\delta x_2;\vec{m} \otimes \vec{m})$.  The relationship
between distribution moments and $q$-point densities is also discussed in 
\cite{bkk}.

\subsection{Covariance-matrix elements}

Given a general measure system $\vec{m}$ defined on $\Omega$ and arguing by analogy with the momentum-variance results above, I can write for any pair of component measures $m_1$ and $m_2$ the quadratic mean
\begin{eqnarray}
M(\delta x) \overline{m_1(\delta x)m_2(\delta x)} &=& M^2\bar m_1\bar m_2 C_2(m_1,m_2;\delta x)
\end{eqnarray}
where $C_2(m_1,m_2;\delta x)$ is the correlation integral up to scale $\delta x$ of the two-point density distribution $\rho_2(m_1,m_2;x_1,x_2)$ on $Q_2$. Note that on the left side $(m_1,m_2)$ are measure values from the same bin in $P_1$ whereas on the right they come from different bins (off-diagonal element of $Q_2$). The corresponding total (co)variance is given by
\begin{eqnarray}
M(\delta x) \sigma^2(m_1,m_2,\delta x) &=& M(\delta x) \, \{\overline{m_1(\delta 
x) \, m_2(\delta x)} - \overline{m_1(\delta x)} \cdot \overline{m_2(\delta x)}\}  \nonumber \\
&\equiv& \Sigma^2(m_1,m_2,\delta x) \nonumber \\
 &=& M^2\bar m_1\bar m_2 \Delta C_2(m_1,m_2;\delta x)
\end{eqnarray}
A comparison of quadratic means at different scales is given by
\begin{eqnarray}
M(\delta x_2)\, \overline{m_1(\delta x_2)m_2(\delta x_2)} - M(\delta x_1)\, 
\overline{m_1(\delta x_1)m_2(\delta x_1)} &=& M^2\bar m_1\bar m_2 C_2(m_1,m_2;\delta x_1,\delta x_2)
\end{eqnarray}
and a (co)variance comparison (generalized form of CLT) is given by
\begin{eqnarray}
 \sigma^2(m_1,m_2,\delta x_2) - N(\delta x_1,\delta x_2) \sigma^2(m_1,m_2,\delta x_1) &=&  \bar{m_1} \cdot \bar{m_2} \, M(\delta x_2) \,  \, \Delta 
C_2(m_1,m_2;\delta x_1,\delta x_2) \nonumber \\
 = \int m_1 m_2 \{ \rho_{2,obj}(\delta x_1,\delta x_2;m_1,m_2) &-& \rho_{2,ref}(\delta x_1,\delta x_2;m_1,m_2) \} dm_1 dm_2  \nonumber \\
 &\equiv& A(\delta x_1,\delta x_2;m_1,m_2),
\end{eqnarray}
or in terms of total (co)variance by
\begin{eqnarray}
\Sigma^2(m_1,m_2,\delta x_2) - \Sigma^2(m_1,m_2,\delta x_1) &=&  M^2\bar m_1\bar m_2 \, \Delta C_2(m_1,m_2;\delta x_1,\delta x_2) \nonumber \\
 &\equiv& {\cal A}(\delta x_1,\delta x_2;m_1,m_2).
\end{eqnarray}

The quantities $\Sigma^2(m_1,m_2,\delta x)$ are elements of a total
covariance  matrix $ {\cal K} (\vec{m},\delta x)$, The per-bin variances
and covariances  $\sigma^2(m_1,m_2,\delta x)$ are elements of the usual
covariance matrix  ${\bf K}(\vec{m},\delta x)$. The quantities 
$\bar m_1 \cdot \bar m_2 \, M(\delta x_2) \, \Delta C_2(m_1,m_2;\delta x_1,\delta x_2)$  are equal to corresponding elements of the integral matrix ${\bf  A}(\delta x_1,\delta x_2;\vec{m} \otimes \vec{m})$, and the quantities $M^2\bar m_1\bar m_2 \Delta C_2(m_1,m_2;\delta x_1,\delta x_2)$ are equal to elements of an integral matrix ${\cal  A}(\delta x_1,\delta x_2;\vec{m} \otimes \vec{m})$, both corresponding to a specific scale interval. We can summarize with the matrix equations
\begin{eqnarray}
{\bf K}(\vec{m},\delta x_2) - N(\delta x_1,\delta x_2) \, {\bf K}(\vec{m},\delta x_1)
 &=& {\bf A}(\delta x_1,\delta x_2;\vec{m} \otimes \vec{m})\end{eqnarray}
and
\begin{eqnarray}
{\cal K}(\vec{m},\delta x_2) -  \, {\cal K}(\vec{m},\delta x_1)
 &=& {\cal A}(\delta x_1,\delta x_2;\vec{m} \otimes \vec{m})\end{eqnarray}
which relate covariance matrix elements and integrals over two-point measure spaces for a bounded scale interval.

\subsection{Covariance matrices in $G_P$ and $G_\Phi$} \label{covar}

We now apply these new concepts to the original problem of understanding the relationship among $A$, $B$ and $\Phi_{p_t}$. The spaces $G_P \equiv G_1(p,\overline{p}n)$ and $G_\Phi \equiv G_1(\overline{n}p/n,\overline{p}n) \approx G_1(p-\overline{p}n,\overline{p}n)$ shown in Fig.  \ref{2d}  provide an archetypal system for the study of extensive/intensive variable combinations and associated  covariance structure.  These spaces are approximately related by a linear shear  transformation ${\bf S}$.  The corresponding covariance matrices ${\bf K}_P$ and ${\bf K}_\Phi$ for an arbitrary distribution common to the two spaces are then defined by the relations
\begin{eqnarray}
{\bf K}_\Phi  &=&
\left( 
\begin{array}[th]{ccc}
\sigma^2_\Phi & \overline{p}\sigma^2_{<\!p\!>N^2} \\
 \overline{p}\sigma^2_{<\!p\!>N^2} &  \overline{p}^2\sigma^2_N \\
\end{array}
\right)  \nonumber \\
&\approx& {\bf \tilde{S}} \cdot {\bf K}_P \cdot {\bf S} \nonumber \\
 &\approx&
\left( 
\begin{array}[th]{ccc}
1 & -1 \\
0 & 1 \\
\end{array}
\right) 
\cdot 
\left( 
\begin{array}[th]{ccc}
\sigma^2_P & \overline{p}\sigma^2_{PN} \\
 \overline{p}\sigma^2_{PN} & \overline{p}^2\sigma^2_N \\
\end{array}
\right) 
\cdot 
\left( 
\begin{array}[th]{ccc}
1 &0 \\
 -1 & 1 \\
\end{array}
\right)  \nonumber \\
 &\approx&
\left( 
\begin{array}[th]{ccc}
\sigma^2_P - 2 \overline{p}\sigma^2_{PN} + \overline{p}^2\sigma^2_N  & 
\overline{p}\sigma^2_{PN} - \overline{p}^2\sigma^2_N\\
 \overline{p}\sigma^2_{PN} - \overline{p}^2\sigma^2_N & 
\overline{p}^2\sigma^2_N \\
\end{array}
\right)
\end{eqnarray} 
These matrix equations compactly represent the relationships among covariance matrix elements derived in Sec. \ref{npt} using a vector representation.

I now define a reference covariance matrix at small (particle) scale in $G_\Phi$ which is analogous to $\vec{\sigma}_R$ in the vector variance formulation and which may serve as a null hypothesis based on a physical model. I obtain the corresponding form in $G_P$ by applying the same shear transformation as before. The reference matrix in $G_\Phi$ has zero covariance and variances $\sigma^2_p$ and $\sigma^2_n$ determined by some unspecified elementary phenomena. For $\delta x_1 \approx a$, $\sigma^2_p$ may correspond to a thermal distribution and $\sigma^2_n \approx 1$ modulo quantum statistics. The momentum $\hat{p}$ represents a characteristic momentum at small scale which may coincide with $\overline{p}$.
\begin{eqnarray}
{\bf K}_{\Phi}(\delta x_1)
&=&  
\left( 
\begin{array}[th]{ccc}
\sigma^2_p & 0 \\
0 & \hat{p}^2\sigma^2_n \\
\end{array}
\right) 
\end{eqnarray}
This choice is particularly simple and could represent an 
independent-particle-emission hypothesis. We can now obtain expressions for covariance difference matrices in both spaces. 
\begin{eqnarray}
{\bf K}_{\Phi}(\delta x_2) 
- N {\bf K}_{\Phi}(\delta x_1)
&=& 
\left( 
\begin{array}[th]{ccc}
\sigma_\Phi^2 - N \sigma^2_p & \hat{p}\sigma^2_{<\!p\!>N^2} \\
 \hat{p}\sigma^2_{<\!p\!>N^2} & \hat{p}^2(\sigma^2_N - N \sigma^2_n) \\
\end{array}
\right) 
\end{eqnarray}
Using the inverse transform
${\bf K}_{P}(\delta x)  =  {\bf \tilde{S}}^{-1} \cdot {\bf K}_{\Phi}(\delta x) \cdot {\bf S}^{-1} $
I can write
\begin{eqnarray} \label{k2}
{\bf K}_{P}(\delta x_2) - N{\bf K}_{P}(\delta x_1) 
 &=&
\left( 
\begin{array}[th]{ccc}
\sigma^2_P - N \sigma_p^2 - \hat{p}^2 N \sigma^2_n &  \hat{p} 
\sigma^2_{PN}  - \hat{p}^2 N \sigma^2_n \\
 \hat{p} 
\sigma^2_{PN}  - \hat{p}^2 N \sigma^2_n & 
\hat{p}^2(\sigma^2_N - N \sigma^2_n) \\
\end{array}
\right)  \nonumber \\
 &=&
\left( 
\begin{array}[th]{ccc}
\sigma^2_\Phi - N \sigma_p^2 +  2 \hat{p} \sigma^2_{<\!p\!>N^2} &  \hat{p} 
\sigma^2_{<\!p\!>N^2}  + \hat{p}^2(\sigma^2_N - N \sigma^2_n) \\
 \hat{p}\sigma_{<\!p\!>N^2} + \hat{p}^2(\sigma^2_N - N \sigma^2_n) & 
\hat{p}^2(\sigma^2_N - N \sigma^2_n) \\
\end{array}
\right) 
\end{eqnarray}
I can now relate covariance matrix elements in the space $G_P = G_1(p,n)$ to corresponding two-point measure integrals in the space $G_2((p,n) \otimes (p,n))$ by the matrix equation
\begin{eqnarray} \label{k1}
{\bf K}(p,n,\delta x_2) - N(\delta x_1,\delta x_2) \, {\bf K}(p,n,\delta 
x_1)
 &=& {\bf A}(\delta x_1,\delta x_2;(p,n) \otimes (p,n)) \nonumber \\
 &=&
\left( 
\begin{array}[th]{ccc}
A_{pp} & \hat{p}A_{pn}  \\
 \hat{p} A_{pn} &  \hat{p}^2 A_{nn}  \\
\end{array}
\right) 
\end{eqnarray}
The LHS is the scale difference between covariance matrices for two-point densities defined on subspaces of $G_1(\vec{m})$ (measure values taken from the {\em same} bin). The RHS represents covariance integrals for two-point densities defined on subspaces of $G_2(\vec{m}\otimes \vec{m})$ (measure values taken from {\em different} bins), the most integral correlation measures that can be defined on $G_2$. Thus, (co)variances of measures defined on single bins in a primary space are interpreted by two-point correlations among measures on the primary space as determined by covariances on pairs of bins.

We now return to the original problem of $A$, $B$ and $\Phi_{p_t}$.
Combining Eqs. (\ref{k2}) and (\ref{k1}) and recalling that $\sigma^2_\Phi - N \sigma_p^2 \approx 2 N \sigma_p \Phi_{p_t}$, and $ 2 \hat{p} \sigma^2_{<\!p\!>N^2} = -B$ we can now relate the upper left elements of both sides
\begin{eqnarray} \label{bterm}
2 N \sigma_p \Phi_{p_t} - B &=& A(a,L;p,p)
\end{eqnarray}
justifying the earlier {\em ad hoc}\/ decomposition of $\Phi_{p_t}$ (Sec. \ref{decomp}) and clarifying its basis. We see that variance sources ($\Phi_{p_t}$) and covariance sources ($B$) may both correspond to net two-point correlations ($A$). $\Phi_{p_t}$ is a `part' of $A$ and not the reverse. This representation only hints at the richness of the system of scale-local correlation measures, even when restricted to pair correlations and (co)variance. Such structures also exist, albeit with increasing algebraic complexity, for arbitrary $q$-tuples and associated higher moments.

\section{Experimental Applications}

The previous material is quite general and is intended to accommodate at least conceptually the transition of any dynamical system across a phase boundary near which the dynamical $DoF$ are ill-defined. However, experimentally we must contend with the hadronic final state of HI collisions and the measures that can be formed there. We now consider some practical measures and analyses, with examples recently proposed in the literature.

\subsection{Available measures and analysis types in the hadronic final state}

In the hadronic final state the basic event-wise measures $\vec{m}$ are particle abundances and momenta, $n_{ij}(\delta x)$ and $p_{ij}(\delta x)$, where $i$ refers to particle species and $j$ indicates a bin index. Special cases include $n_{ij}(a) = 1$, $n_{ij}(L) = N_i$ -- event abundance for species $i$, $p_{ij}(a) = $ particle momentum and $p_{ij}(L) = P_i$ -- event total momentum for species $i$. These constitute the primary measures in $\vec{m} = \{m_{ij}\}$. We can also add an integer index $n$ randomly assigned to particles.

A secondary measure system is derived from correlation analysis of individual events. Distribution statistics ($<\!p_t\!>$) and model parameters (slope parameter) are simple examples. HBT correlation analysis in principle provides event-wise estimates of source radii, flow analysis can estimate the reaction plane and azimuthal anisotropies, and more general event-wise correlation analysis can  provide a complete topological assessment of each event in terms of information or dimension transport. A minimal set of correlation parameters which completely characterizes the event can then be included in $\vec{m}$.

Depending on the abundance of a given particle species one can pursue an integral or differential analysis. In the case of limited abundance one can compare total or per-bin covariance matrix elements at two scales, element by element. These are just central-limit tests as represented by Eq. (\ref{k1}).
In the case of greater abundance one can examine directly the integrands of elements of ${\bf A}(\delta x_1,\delta x_2;\vec{m}\otimes \vec{m})$. One can analyze differences $\rho_{2,obj}(\delta x_1,\delta x_2;m_1,m_2) - \rho_{2,ref}(\delta x_1,\delta x_2;m_1,m_2)$ or preferably density ratios $\rho_{2,obj}/\rho_{2,ref}$ to search for departures from a reference.

\subsection{Partition systems}

The partition systems on $P$ space directly accessible in the hadronic final state are at the particle ($a$) and event ($L$) scales. Additional partititions can only be defined on $\{\vec{m},n\}$ and therefore only indirectly qualify as partitions on scale, to the extent that there are correlations between elements of $\{\vec{m},n\}$ and $P$ space. For instance, strong correlation between longitudinal momentum (rapidity) and axial position means that a partition on rapidity or pseudorapidity provides some ability to partition $P$ space at different scales on axial position. However the scale interval is limited because this correlation is not perfect due to thermal and other fluctuations in longitudinal momentum about the basic Bjorken expansion correlation. Results obtained with such a partition must be carefully interpreted over a restricted scale interval. 

A partition on azimuth angle is also possible, but its interpretation depends on the degree of correlation between azimuth angle in momentum space and configuration space. Depending on the emission model this could be very stong (emission from an opaque source) or negligible (emission from a transparent source). Partition of the random index $n$ is the most indirect but also least biased partition method, but its relationship to a scaled partition is not clear in general.

\subsection{Nonstatistical fluctuations}

The distinction between nonstatistical (`dynamical') and statistical fluctuations or variance can be seen as insubstantial given the discussion of scale-dependent variance presented above. `Statistical' variance is accumulated below and within some scale interval characteristic of a correlation onset, or below the scale interval of interest. Dynamical fluctuations as discussed here simply correspond to variance accumulated on a complementary (larger) scale interval -- the one of immediate interest. In a different analysis some statistical fluctuations might be considered dynamical by the same argument. For discussion purposes we adopt a distinction of convenience between statistical variance and its complement on scale.

The term $A$ in Eq. (\ref{bterm}), which represents the most global aspect of nonstatistical fluctuations, can be decomposed into several contributions as follows
\begin{eqnarray} \label{multdep}
 A(a,L;p,p)&\approx& \{ \bar N \,(\hat{q}_{res} - 1 + \sigma^2_{\hat{q}_{res}} )  +  \bar N(\bar N - 1) \,(\hat{q}_{qc} - 1 + \sigma^2_{\hat{q}_{qc}} ) \}  \, \sigma^2_{p_t} \nonumber \\
&+&  \bar N(\bar N - 1)\, \sigma^2_{\bar p_t} + \cdots
\end{eqnarray}
where $\hat{q}_{res} \in [1,2+]$ represents the effective sibling number for secondary particles derived from a single primary parent ({\em e.g.,} resonance decays), $\hat{q}_{qc} \in [0,2]$ represents the effective sibling number for two-point quantum correlations (HBT and FSI), $\sigma^2_{\hat{q}_{res}}$ represents event-wise fluctuations in ${\hat{q}_{res}}$, $\sigma^2_{\hat{q}_{qc}}$ relates similiarly to $\hat{q}_{qc}$ and $\sigma^2_{\bar p_t}$ represents event-wise variation of the parent mean ({\em e.g.,} distortion of the parent $p_t$ distribution by event-wise fluctuating DCC or radial flow contributions). Thus, possible EbyE fluctuations in single-point {\em and} two-point aspects of the parent distribution are here included with possible sources of two-point correlation as contributions to $A$. Mean values of quadratic forms are essentially the only aspects of EbyE variation of the parent distrbution accessible to a two-point or variance analysis. Note that an uncorrelated system ($\hat{q} \equiv 1$) and stationary parent ($\sigma^2 \equiv 0$) result in $A \rightarrow 0$, consistent with the CLT. This gives some impression of the complex task of interpreting a nonzero {\em or} zero value for $A$. Some unfolding of these possibly bipolar contributions can be achieved by noting multiplicity dependence, but there is not a unique inversion in general.

\subsection{The subevent method}

In the subevent method \cite{vol} each event is partitioned into two parts at large scale ($\approx L/2$), for instance by a pair of disjoint bins on rapidity, azimuth angle or a random particle index $n$. One then extracts the covariance of one or a pair of additive measures $m$ $(= n(L/2),p(L/2)\cdots$) or ratios $<\!m\!> \equiv m_1/m_2$ obtained from the two parts. Effectively, the two-point density analyzed is $\rho(L/2,L;m_1,m_2)$ on $G_2$. We can interpret this method within the present context by referring to Eq. (\ref{adef}). The LHS is the variance difference referred to in \cite{vol} as the `direct' method (in contrast to the subevent method). For specific cases (scale intervals) the RHS is the same covariance which forms the basis for the subevent method.  For these cases the quantity $A(\delta x_1,\delta x_2;x_1,x_2)$ can be identified with the quantity $\sigma^2_{x,dyn}$ in \cite{vol}, thus also providing the relationship between $\Phi_{p_t}$ and the subevent method. (The discussion of $\Phi_{p_t}$ in \cite{vol} assumes lack of correlation between $<\!p_t\!>$ and $N$, so the corresponding covariance term $B$ in Eq. (\ref{bterm}) is missing.)

One must carefully specify the scale interval being used for a given analysis. The subevent method as defined refers approximately to the scale interval $[L/2,L]$, whereas the `direct' method as discussed in \cite{vol} and the $\Phi_{p_t}$ measure as it is usually presented refer to the inteval $[a,L]$. The subevent method should therfore nominally be sensitive only to correlation developed or total variance accumulated in the interval $[L/2,L]$, and should not be directly comparable with an analysis based on $[a,L]$. However, because the partition methods employed are not applied directly to the primary space but only indirectly to a measure space (momentum space or a random index) it is possible for small-scale correlations to be included at some unknown level, prompting a recommendation to partition momentum space at large scale in such a way as to exclude issues of two-track resolution, BE/FD correlations and FSI. One effectively cuts on density $\rho(a,L;p_1,p_2)$ in $G_2(p_1,p_2)$ so as to exclude a neighborhood around the main diagonal (at small scale) which contains quantum correlations and final-state interactions. 

Various multiplicity dependencies are considered in \cite{vol} as means to unravel multiple contributions to $\sigma^2_{x,dyn}$ similar to the contributions included in Eq. (\ref{multdep}).  Missing from \cite{vol} are possible contributions represented by terms $\sigma^2_{\hat{q}}$ for resonances and quantum correlations from event-wise variation of the {\em two-point} parent distribution. As with all global-variables analyses direct analysis of two-point density ratios on space $G_2(m_1,m_2)$ provides a more complete way to unfold differing contributions to variance comparison measures when the data have sufficient statistical power to support this approach.

The subevent method is analogous to forming a mixed-pair reference from similar events, execept that the `events' compared are subevent components of a single event. Use of large-scale event partitions to establish a statistical reference may be difficult to interpret in some cases, albeit computationally convenient. Assumptions must be made about the correlation structure of events which may be difficult to justify. An alternative for the study of large-scale correlations may be to compare sibling {\em and mixed} subevents -- subevent pairs formed in similar ways from the same event and from similar events. This procedure would form part of a general scale-local approach to two-point correlations spanning the scale interval from particle pairs to subevent pairs, with the same general analysis method used throughout.

\subsection{Ratio fluctuations and resonance production}

In \cite{mul} it is proposed to study event-wise fluctuations in hadron multiplicity ratios in order to determine resonance abundances at chemical freezeout. The method is based on a change in the fluctuation structure of $+/-$ multiplicity ratios to the extent that resonance decays (R) as well as thermal production (T) contribute to hadron yields. If a $+/-$ multiplicity ratio is defined by $R_{12} =  N_1 /  N_2$ and we assume no correlations among resonances ($\sigma^2_r \approx \bar N_r$) then the relative variance in the ratio is given by ($r$ is a resonance type index)
\begin{eqnarray} \label{r12}
{\sigma^2_{R_{12}} \over \bar R_{12}^2} &=& \left\{ {g_1 \over \bar N_1} + { g_2 \over \bar N_2} - 2 \, {g_{12} \, \over \sqrt{\bar N_1 \bar N_2}}   \right \}
\end{eqnarray}
where
\begin{eqnarray}
g_1 &=& {1 \over \bar N_1} \{ \sigma^2_{T,1} + \sum_r \bar N_r \, \overline{n^2_{r,1}} \} \nonumber \\
g_2 &=& {1 \over \bar N_2} \{ \sigma^2_{T,2} + \sum_r \bar N_r \, \overline{n^2_{r,2}} \} \nonumber \\
g_{12} &=& {1 \over \sqrt{\bar N_1 \bar N_2}} \{ \sigma^2_{T,12} + \sum_r \bar N_r \, \overline{n_{r,1} \, n_{r,2}} \} 
\end{eqnarray}
If we further assume that $\bar N_1 \approx \bar N_2 \approx \bar N = \bar N_T + \bar N_R$, $\bar N_R \equiv \sum_r \bar N_r $, $\alpha_R \equiv \bar N_R / \bar N$, $\alpha_T \equiv 1 - \alpha_R$, $ \sigma^2_{T,i}\approx  \bar N_T$, $\sigma^2_{T,12} \approx 0$, $n_{ri} \equiv 1$, then we have $g_1 \approx g_2 \approx 1$ and $ g_{12} \approx \alpha_R$. Therefore, $\sigma^2_{R_{12},obj} / \sigma^2_{R_{12},ref} \approx 1 - \alpha_R$. These assumptions apply to correlation-free thermal pion emission combined with a single thermal resonance type decaying strictly to $\pi^+ - \pi^-$ pairs as a simple example.

Eq. (\ref{r12}) for $G_1(n_1,n_2)$ is a variant of Eq. (\ref{sigphi}) for $G_1(p,n)$ (equivalent in form if the latter is divided through by $\bar P$). There are two similar cases described by this algebra: distributions on $(n_1/n_2,n_1 \cdot n_2)$ and on $(n_1 - n_2, n_1 + n_2)$. The relative fluctuation distribution is essentially the same in the two spaces if a gaussian model and linearization are appropriate. The $obj/ref$ ratio for multiplicity {\em products} (or sums) in the same case would be $\sigma^2_{\Pi_{12},obj} / \sigma^2_{\Pi_{12},ref} \approx 1 + \alpha_R$ \cite{edw}, where $\Pi_{12} \equiv N_1 \cdot N_2$. For the simple example considered, the thermal contribution is uncorrelated on $G_1(n_1,n_2)$, with equal variance on all variable combinations, whereas the resonance contribution is perfectly correlated along $N_1 = N_2$. A superposition of these two contributions with fractions $\alpha_T$ and $\alpha_R$ gives the stated result.

While correlations in the resonance contribution decrease the variance of the difference and ratio distributions they {\em increase} the variance of the marginal distributions on $N_1$ and $N_2$ and the sum and product distributions. The variance of $<\!p_t\!> = P_t/N$ is also increased by the presence of resoncances because the effecive sample number $N_{eff}$ is {\em reduced} by correlations. In the limit of all resonances decaying to pion pairs and no thermal pions the observed number of particles is twice the number of statistical $DoF$ as reflected in an effective sibling index $\hat{q}_{res} \approx 2$ in Eq. (\ref{multdep}), leading to a doubling of observed $<\!p_t\!>$ variance over the CLT expectation for the observed particle multiplicity. 

By some combination of fluctuation measurements of ratios/products or sums/differences and invocation of known resonance-decay schemes one can hope to infer the fractional resonance yield at chemical freezeout. Precision studies of resonance production, which are interesting in themselves and essential as a baseline to understand other aspects of the hadronic final state, require the formation of a reference system, so that one is dealing not with the ratio/product or sum/difference covariance system for real events alone but rather with differential departures from a reference. There is also the problem that other correlation sources may compete with resonance decay to alter $N_+/N_-$ fluctuations.

\subsection{Covariance analysis of global variables}

In \cite{gav} it is proposed to increase sensitivity to possible QGP `signals' by a combined analysis of the covariance of two or more global event properties. Event-wise antiproton multiplicity and transverse emitter radius are considered as an example. This approach is similar to the analysis format presented in Sec. \ref{covar}, but without use of a reference to form a comparison measure. The centrality dependence of covariance is itself taken as the indicator of unexpected behavior. One interesting aspect of this proposed application is that one of the measures is not a simple disttribution statistic like total multiplicity or mean $p_t$ but a more complex correlation aspect of the two-point momentum disttribution. This proposal is consistent with a general program to measure fluctuations of all kinds in the neighborhood of a phase boundary in order to better understand the underlying Lagrangian by enumerating and identifying the relevant $DoF$ for various constraints.

The underlying model premise in \cite{gav} is that, depending on centrality, collisions may fluctuate between two disconnected event classes, one ($q$) manifesting characteristics associated with QGP formation ({\em e.g.,} increased emitter radius and antiproton abundance mean values) and the other ($h$) manifesting conventional hadronic systematic variation with participant number. The measure system in this case is $(m_1,m_2) = (N_{\bar p}, R_t)$, and the centrality dependence of the covariance is contrasted with that of the mean values of  the measures for sensitivity in terms of this fluctuation model. While the paper emphasizes covariance measures it admits that variances would exhibit similar centrality dependence for this model. The central point is that attention to fluctuation measures is critical for detection of intermittent phenomena in collisions.

The basic elements of the simulation are the estimated  baseline measure fluctuations in the hadronic and QGP regimes, and the assumed shifts in mean values from one event class to the other. These assumptions determine all relevant aspects of the simulation. Mean values for the two event classes [h,q] are given by $\bar N_{\bar p}(b) \approx [56,78]\,(N_{ch}(b)/N_{max})$ (antiproton multiplicity into STAR acceptance) and $\bar R_t(b) \approx [6,9 ~ fm] \, (N_{ch}(b)/N_{max})^{1/2}$, where $N_{ch}$ is the charged-particle total multiplicity used to infer event centrality and $N_{max}$ is the mean value for central collisions. Variances are approximated by $\sigma^2_{N_{\bar p}} \approx \bar N_{\bar p}$ and $\sigma^2_{R_t} \approx \bar R^2_t$, the second dominated by an estimate of measurement error rather than intrinsic statistical fluctuations. The result is a rather large increase in covariance in a crossover region of centrality where the two event types are nearly equiprobable. We observe in a scatter plot on $(N_{\bar p},R_t)$ that the joint distribution is bimodal due to the chosen relationship between assumed mean-value shifts and variances.

The assumption of catastrophic fluctuations between two event types maximizes the effect of the model on variances and covariance. An alterative scenario consisting of a partial transition (still first order) over some fraction of the event volume (having its own fluctuation structure) would produce a significantly smaller effect on correlation measures. These model assumptions do however correspond to some popular QGP scenarios. We note that an expected major source of covariance, collision geometry and stopping fluctuations (`volume' fluctuations) as illustrated in Fig. \ref{2d} and present for either event type, is here probabably underestimated by the Poisson variance of $N_{ch}$. 

Having modelled the most dramatic QGP scenarios one might then consider how to establish by these correlation measures either that a QGP hypothesis is falsified at some level, or that some fraction of some events does indeed undergo a transition to a nonhadronic phase. In the adopted scenario the input to the corelation measures is bimodal, a rather strong statistical manifestation. A weaker and possibly  more realistic situation to simulate is one in which anomalous events form a tail structure on a hadronic core. In this case one has to compare carefully the centrality dependence of covariance matrix elements to that for a reference and perform a significance study of the result. 

There is much technical work needed to elaborate these and other EbyE analysis methods. We see  from these examples and interpretations that a general description of scale-local covariances and comparison measures permits a comprehensive interpretation of some apparently dissimilar methods.

\section{Toward a Generalized Central Limit Theorem}

From the original concept of the elementary central limit theorem as an integral variance comparison measure for a gaussian fluctuation model, we have enlarged the analysis to multiple measures described by a covariance matrix, have expressed the basic comparison in a scale-dependent manner, have espressed the source of departure from the CLT in terms of a matrix of scale integrals comparing object and reference two-point densities and have considered as the most differential form of comparison a direct analysis of density ratios on two-point measure spaces.

From the vantage point of this scale-local picture we can now generalize the central limit theorem and reinterpret its role in the description of many-body systems. In terms of the new concepts the elementary CLT as presented in Sec. \ref{clt} compares total variance for a single measure at two points on scale to determine whether additional variance has accumulated over the interval. The CLT states that if there is no net correlation introduced within a scale interval the total variance of a measure is the same at each endpoint -- $\Sigma^2(p,p,\delta x_2) - \Sigma^2(p,p,\delta x_1) = 0$. 

We have subsequently developed a greatly extended version of this elementary comparison procedure and generalized it to a pair-wise total (co)variance comparison for an arbitrary measure system as represented by
\begin{eqnarray}
 {\cal K} (\vec{m},\delta x_2) - {\cal K} (\vec{m},\delta x_1) &=&  {\cal A}(\delta x_1,\delta x_2;\vec{m}\otimes \vec{m})
\end{eqnarray}
and have given the general form of the source of any total variance change in terms of a matrix of covariance integrals over two-point measure densities represented by ${\cal A}$. We can therefore generalize the CLT to a statement about the scale 
invariance of covariance:
\begin{quote}
The total covariance matrix ${\cal K} (\vec{m},\delta x)$ for a measure system on the space $\Omega$ is {\em scale invariant}\/ over a bounded scale interval $[\delta x_1,\delta x_2]$ in the absence of all categories of {\em net}\/ two-point measure correlation over the same scale interval. Departure from scale invariance is represented by covariance integrals of density differences on corresponding two-point measure spaces.
\end{quote}
In this formulation we have represented the possible sources of accumulated covariance as integrals over two-point measure spaces. We have also introduced the possibility that a uniform reference (implicit in standard variance calculations) may be replaced by an arbitrary reference or model appropriate to a given problem. The resulting system offers the opportunity to choose from a continuum of techniques and references, depending on the fluctuation model and statistical power of the data.

\section{Conclusions}

We have as our nominal goal identification and study of a quark-gluon plasma,  a simplistic terminology for a complex phenomenon. It may be that the emergence of a color-deconfined QCD system is subtle and topological rather than catastrophic and thermodynamic. Production of a QCD plasma as a catastrophic process with discontinuous signals may not be forthcoming. To insure the best opportunity for discovery we require sensitive quantitative comparison measures which extract complete correlation information from collision events.

In studying the correlation structure of HI collision events at the CERN SPS we have been struck by their sameness and ordinariness, the apparent close adherence to simple statistical fluctuations in the distribution of mean $p_t$ and the wide-ranging description of hadronic ratios and abundances by a simple statistical model with a universal chemical freezeout temperature. These two results are deeply connected by their relationship to the central limit theorem and related maximum symmetry principle. A common supposition has been promoted that QCD plasma formation at high energy densities may lead to dramatic, or easily quantifiable, changes in the correlation structure of the multiparticle final state, changes detectable by elementary correlation measures. Given the unspectacular SPS results we are now forced to re-evaluate our need for sensitivity.

A central goal in EbyE analysis has been to develop global comparison measures sensitive to excess variance which might signal residual correlations due to incomplete equilibration -- possibly structure remaining from a phase transition. We have seen that these global measures are ultimately based on the central limit theorem as a statistical reference, and the CLT is in turn based on a scale-integral concept. The present treatment makes it clear that global-variables analysis based on scale integrals -- an unavoidable strategy in case of low statistical power -- is rather limited in its sensitivity, interpretability and power to discriminate among various correlation sources. 

Statistical power (event multiplicity) determines the optimum scale interval and scale resolution of a correlation analysis. Low statistical power requires the use of integrals over space and scale to extract statistically meaningful correlation measures, with accompanying loss of information. Higher statistical power makes possible extension of the analysis to scale-local differential measures and direct analysis of densities on two-point measure spaces. One can then study the scale dependence of density ratios to extract full correlation information from data. Future developments include direct determination of two-point densities, analysis of scale-local topological measures extracted from these densities and extention of the analysis to higher moments ($q > 2$).

The question of scale invariance represented by the CLT is deeply related
to the nature of distribution symmetries. Scale-invariance is equivalent
to constant autocorrelation -- Eq. (\ref{deq}) -- which is in turn an indication of maximum symmetry. Thus, the CLT in its generalized  form is a statement about
symmetry variation over a bounded scale  interval, {\em a potentially new
insight}. With a uniform distribution as reference one can inquire
whether an object distribution is maximally symmetric over a scale
interval and how its symmetries evolve dynamically.  More generally, one can ask whether object and reference
distributions have the {\em same}\/ symmetry or autocorrelation over some
scale interval. This presents a new approach to quantitative description of
nonequilibrium systems and the equilibration process.

Scale-dependence of symmetry variation is the underlying issue in the study of phase transitions. Near the QCD phase boundary hadrons and partons as dynamical objects are transitory and ill defined. Basing a dynamical description on a particle hypothesis near a phase boundary is a poor strategy. To provide a fundamental treatment one must ultimately adopt a descriptive system in which particles are not fundamental elements, in which arbitrarily correlated measure systems play the central role.  A generalized central limit theorem provides a basis for such an approach. One can study the symmetry dynamics of a measure system relative to a maximally symmetric system or other symmetry model. 

In a broader sense we are probing the dynamics of vacuum symmetries
within  a violent collision. The more traditional study of symmetry {\em
statics}\/  is carried out in asymptotic regions of a constraint space far
from any phase boundary. Symmetry  {\em dynamics}\/ is important in the
neighborhood of a phase boundary. The QCD phase boundary marks a region of
constraint space where the nature of vacuum symmetries is changing dramatically and fundamentally.  This is a challenging undertaking requiring the best possible analysis techniques.

\section{Acknowledgments}

I appreciate helpful discussions with or useful comments from M. Ga\'{z}dzicki,  A.J. Poskanzer, K. Rajagopal, J.G. Reid, G. Roland, E. Shuryak, S. Voloshin and D.D. Weerasundara.
Some of this work was initiated with the support of RHIC
R\&D grant DE-FG06-90ER40537.  I am thankful for continuing research
support from the USDOE under grant DE-FG03-97ER41020.

\end{document}